\documentclass{aa}  

\usepackage{graphicx}
\usepackage{amsmath}
\usepackage{amssymb}
\usepackage{txfonts}
\usepackage{natbib}
\usepackage[]{hyperref} 
\usepackage{longtable}
\usepackage{textcomp,gensymb}
\usepackage[normalem]{ulem}
\usepackage{booktabs}
\usepackage[usenames, dvipsnames]{xcolor}

\def\kms{{\rm km\,s^{-1}}}

\begin{document}

   \title{Riding the kinematic waves in the Milky Way disk with Gaia}


   \author{P. Ramos
   			,
          T. Antoja
          \and
          F. Figueras
          }

   \institute{Institut de Ciències del Cosmos, Universitat de Barcelon  (IEEC-UB), Martí i Franquès 1, 08028 Barcelona, Spain\\
              \email{pramos@fqa.ub.edu}
             }

   \date{Received date; accepted date}

 
  \abstract
   {\textit{Gaia} DR2 has delivered full-sky 6-D measurements for millions of stars, and the quest to understand the dynamics of our Galaxy has entered a new phase.}
   {Our aim is to reveal and characterize the kinematic sub-structure of the different Galactic neighbourhoods, to form a picture of their spatial evolution that can be used to infer the Galactic potential, its evolution and its components.}
   {We take $\sim$5 million stars in the Galactic disk from the \textit{Gaia} DR2 catalogue and build the velocity distribution of many different Galactic Neighbourhoods distributed along 5 kpc in Galactic radius and azimuth. We decompose their distribution of stars in the $V_R$-$V_\phi$ plane with the wavelet transformation and asses the statistical significance of the structures found.}
   {We detect many kinematic sub-structures (arches and more rounded groups) that diminish their azimuthal velocity as a function of Galactic radius in a continuous way, connecting volumes up to 3 kpc apart in some cases. The decrease rate is, on average, of $\sim$23 $\kms$kpc$^{-1}$. In azimuth, the kinematic sub-structures present much smaller variations. We also observe a duality in their behaviour: some conserve their vertical angular momentum with radius (e.g., Hercules), while some seem to have nearly constant kinetic energy (e.g., Sirius). These two trends are consistent with the approximate predictions of resonances and of phase mixing, respectively. Besides, the overall spatial evolution of Hercules is consistent with being related to the Outer Lindblad Resonance of the Bar. 
   We also detect structures without apparent counterpart in the vicinity of the Sun. }
   {The various trends observed and their continuity with radius and azimuth allows for future work to deeply explore the parameter space of the models. Also, the characterization of extrasolar moving groups opens the opportunity to expand our understanding of the Galaxy beyond the Solar Neighbourhood.}


   \maketitle
%

\section{Introduction}
\label{sec:introduction}

The velocity distribution of the stars in our Galaxy has been deeply studied during the last decades \citep[e.g.][]{Dehnen1998b,Skuljan1999,Famaey2005,Antoja2008}, in an attempt to understand the Galactic potential and its components. By comparing this distribution with models of the orbital effects of the Galactic bar and the spiral arms, which cause sub-structure and gaps in the velocity distribution, it became clear that only the exploration of regions outside the Solar Neighbourhood would yield a proper constraint on the potential  \citep{Bovy2010b,Antoja2011,Quillen2011,McMillan2013}.

The first studies that investigated Galactic regions relatively distant from the Sun, even with the limitations in the number of stars and precision of their samples, have already presented relevant results. \citet{Antoja2012} was the first study showing that the kinematic sub-structures in the Solar Neighbourhood change their velocities when observed in other volumes of the RAVE data \citep{Steinmetz2006}. In a follow-up of that work, \citet{Antoja2014} demonstrated that the azimuthal velocity of the Hercules stream decreases with Galactic radius consistently with the effects of the Outer Lindblad Resonance of the Galactic bar, as proposed by the models of \citet{Dehnen2000} and \citet{Fux2001}. With data from RAVE, LAMOST \citep{Cui2012} and APOGEE \citep{Majewski2017}, later studies have also shown that the kinematic groups are position dependent \citep{Xia2015, Liang2017,Monari2017,Kushniruk2017, Quillen2018, Hunt2018, Monari2018}.


Now, the \textit{Gaia} mission \citep{gaiamission}, which allowed astrophysicist to produce valuable science with only a year of observation time (\textit{Gaia} DR1, \citealt{dr1}), has the chance to show its unparalleled scientific value once more with the second data release (\textit{Gaia} DR2, \citealt{dr2}). After 668 days of surveying, the unprecedented quantity, quality and extension of the \textit{Gaia} data has already uncovered a new configuration of the velocity distribution in the Solar Neighbourhood with stars organized in multiple thin arches never seen before \citep{Katz2018a} as well as the beautiful continuity of these sub-structures through the Galactic disk \citep{Antoja2018}. All these new findings make us reconsider all the hypotheses behind the existence of kinematic sub-structure in the disk, which requires a more detailed characterization of the velocity distribution.

In this work, we use $\sim$5 million sources in \textit{Gaia} DR2 with positions, radial velocity, proper motions and parallaxes to study  the  known and the new main features of the velocity distribution of the Solar Neighbourhood, as well as to examine the most remote regions in the disk explored thus far. For this, we use the Wavelet Transform to detect overdensities in the velocity distributions and draw their evolution with Galactic radius and azimuth. 

This paper is organized as follows. In Section \ref{sec:sample}, we describe the observational data used and its partition into sub-samples. Section \ref{sec:wavelets} presents the formalism of the Wavelet transformation and the methodology to detect structures. Section \ref{sec:solar} characterizes the kinematic distribution of the Solar Neighbourhood and Sect. \ref{sec:vicinity} explores the evolution of these at different Galactic neighbourhoods. Then, in Sect. \ref{sec:discussion} we discuss the implications of our findings and propose possibles future lines of research. Finally, we present the main conclusions of this work in Sect. \ref{sec:conclusions}. 


\section{Data and sample selection}
\label{sec:sample}

\textit{Gaia} is an all-sky astrometric satellite from which we can now obtain position in the sky ($\alpha$, $\delta$), parallax ($\varpi$) and proper motions ($\mu_{\alpha*}$, $\mu_\delta$), along with the estimated uncertainties ($\sigma$) and correlations, for $\sim$1.3 billion sources. Also, line of sight velocities ($v_{los}$) are available for >7.2 millions stars with an effective temperature between 6900K and 3550K \citep{Katz2018b}, corresponding roughly to spectral types from F2 to M2 \citep{Pecaut2013}. 

\begin{figure}
\centering
\includegraphics[width=\linewidth]{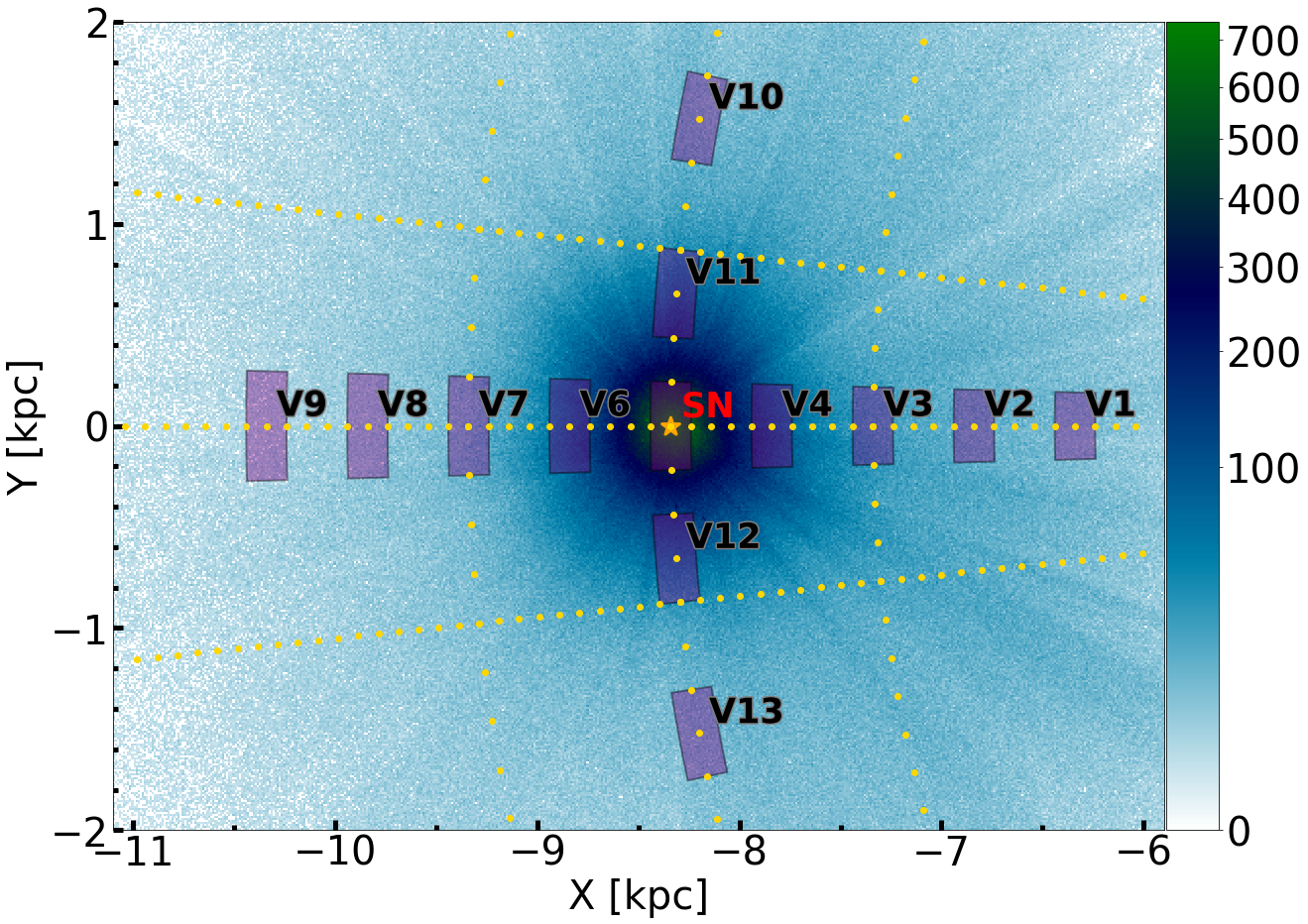}
\caption{Distribution of the sample in configuration space. We show the histogram of the 5,136,533 stars (see text) with a binning of 10x10 pc. The Sun is shown as an orange star and the Galactic Centre is located at $(X,Y)=(0,0)$. The golden dots indicate the centres of the volumes used for the exploration at different Galactic neighbourhoods. Each radial line contains 51 volumes, and each arch at fixed radius has 21 volumes. The purple patches correspond to the sub-samples from the wider grid for which we perform a detailed study of their velocity distribution. The properties of these sub-samples are summarized in Table \ref{tab:sectors}.}
\label{fig:sample}
\end{figure}

Our goal to study the kinematic structure required all 6 phase-space coordinates, which means that our sample consists only of stars with observed radial velocity. On top of that, we limited ourselves to sources with "good" parallax ($\varpi/\sigma_\varpi >5$) in order to use the estimator $1/\varpi$ as distance (this choice and its effects are discussed in Appendix \ref{App:parallax}). Additionally, we focus on disk stars by selecting the heights (Z) between $\pm$0.5 kpc, which corresponds roughly to the 10- and 90-percentile of the whole sample. As a result, our sample is composed of 5,136,533 stars. A significant fraction of Red Giants over Dwarfs for the most distant sub-samples is expected, with a more homogeneous mix near the Sun's position. We note, however, that the youngest stars are missing due to the restriction in effective temperature.

For simplicity, we adopted a Cylindrical Galactocentric coordinate system  since it is better suited for systems with rotational symmetry, as is roughly the case for the Milky Way (MW) disk. Thus, we fixed the reference at the Galactic Centre (GC) with the radial direction ($R$) pointing outwards from it, the azimuthal ($\phi$) negative in the direction of rotation, and the vertical component ($Z$) positive towards the North Galactic Pole. In this scenario, we took the Sun to be at $R_\odot = 8.34$ kpc \citep{Reid2014}, $\phi_\odot = 0^{\circ}$ and $Z_\odot = 14$ pc \citep{Binney1997}. Following from the choice of coordinate system, we studied the stars in the $\dot{R}=V_R$, $R\dot{\phi}=V_\phi$ velocity plane. To compute these from the \textit{Gaia} observables, we adopted a circular velocity for the Sun of $240$ $\kms$ as in \cite{Reid2014}, and a peculiar velocity with respect to the Local Standard of Rest (LSR) in Cartesian coordinates of $(U_{\odot},V_{\odot},W_{\odot})=(11.1, 12.24, 7.25)$ $\kms$ from \cite{Schonrich2010}. The uncertainty in velocities was computed by error propagation with the Jacobian matrix of the transformation, including the correlations, from ICRS to Galactocentric Cylindrical coordinates.

Figure \ref{fig:sample} displays the distribution of the sample in the disk plane, highlighting the isotropy of the \textit{Gaia} data except for the radial features related to extinction. Superposed, we plot the 13 sub-samples that we used to study the variation of the velocity distribution with radius and azimuth. The properties of these sub-samples are shown in Table \ref{tab:sectors}. The Solar Neighbourhood (SN) has the largest number of stars and the smallest uncertainties, allowing for a detailed study of the kinematic structures. Then V1 to V9, along \textit{R}, let us explore the changes with Galactocentric distance and the effects of the sample size. Finally, we used V10 to V13, located at mirrored azimuths with respect to the Sun, to investigate variations in azimuth. All these volumes are part of a wider grid of sub-samples, whose centres are represented by the golden dots in Fig.\ref{fig:sample}. Their sizes are kept constant, with 200 pc in radius and 3 deg in azimuth, resulting in an area of $\sim$0.18 kpc$^2$ (smaller towards the GC and larger in the opposite direction). Overall, we explored the Galaxy disk with 216 volumes divided into three sweeps along \textit{R} (at $\phi=\{-6,0,6\}$ deg.) and three along $\phi$ (at $R=\{R_\odot-1,R_\odot,R_\odot+1\}$ kpc). To produce smooth results, we had to balance the step size and overlapping of the volumes with the range of exploration: we decided to layout the centres every 100 pc between 6.04 and 11.04 kpc, and every 1.5 degrees between -15 and 15 degrees. This translates into an overlapping such that each area is covered completely by the half of the two adjacent areas. 

\begin{table}
\centering
\begin{tabular}{lccccc}
\hline\hline
Volume &R &$\phi$ &Counts&$\tilde{\sigma}_{vel}$ \\
 &[kpc]&[º]& &[km/s]\\
\hline
SN & 8.24-8.44 & -1.5$\sim$1.5 &  435,801 & 0.83\\
V1 & 6.24-6.44 & -1.5$\sim$1.5  & 6,920 & 3.45\\
V2 & 6.74-6.94 & -1.5$\sim$1.5 &  10,959 & 2.53\\
V3 & 7.24-7.44 & -1.5$\sim$1.5 & 20,743 & 1.93\\
V4 & 7.74-7.94 & -1.5$\sim$1.5 &  88,184 & 1.99\\
V6 & 8.74-8.94 & -1.5$\sim$1.5 & 77,457 & 1.95\\
V7 & 9.24-9.44 & -1.5$\sim$1.5 & 13,006 & 1.94\\
V8 & 9.74-9.94 & -1.5$\sim$1.5 & 5,994 & 2.39\\
V9 & 10.24-10.44 & -1.5$\sim$1.5 & 3,340 & 2.79\\
V10 & 8.24-8.44 & -12$\sim$-9 & 10,579 & 1.63\\
V11 & 8.24-8.44 & -6$\sim$-3 & 72,524 & 1.75\\
V12 & 8.24-8.44 & 3$\sim$6 & 64,849 & 1.71\\
V13 & 8.24-8.44 & 9$\sim$12 & 9,753 & 1.62\\
\hline
\hline
\end{tabular}
\caption{Properties of the main sub-samples in our study. In the first column, the name of the volume is shown accordingly to Fig.\ref{fig:sample}. Columns 2 and 3 contain the ranges of the radial and azimuthal coordinates. The last two columns show the summary statistics \textit{Counts} and $\tilde{\sigma}_{vel}$ that are, respectively, the number of stars in the sub-sample and the square root of the trace of the median velocity error covariance matrix in Cylindrical coordinates.}
\label{tab:sectors}
\end{table}


\section{Methods}
\label{sec:wavelets}

In order to detect sub-structures in the velocity plane ($V_R$, $V_\phi$), we used the Stationary Wavelet Transform (WT, \citealt{Starck2002}). Particularly, we used the \textit{à trous} algorithm, which allows the treatment of discrete data, like time series or images, without reducing the number of pixels (in our case, a 2D histogram). 
The WT decomposes the signal into several layers, or scales, of the same dimensions as the original, each comprising a range of frequencies (in case of time) or 
sizes (in the case of space). In practice, the WT returns a map of coefficients for every scale such that zero means that the signal is constant in a neighbourhood whose size is determined by the scale (as we shall see below), whereas positive coefficients correspond to overdensities and negative ones are related to underdensities. The algorithm works by successive application of smoothing filters that, when subtracting one from the previous, results in the mentioned maps of wavelet coefficients. A more detailed explanation can be found in \cite{Antoja2008, Antoja2015b,Kushniruk2017} and references therein, where the WT is applied to similar and other science cases. 
Here we used the Multiresolution Analysis (MRA) Software, developed by CEA (Saclay, France) and Nice Observatory \citep{Starck1998b}.

By construction, in the \textit{à trous} algorithm structures found at scale \textit{j} have sizes approximately between  $2^j$ and $2^{j+1}$ pixels. The conversion from pixels to physical units ($\kms$ in this case) is given simply by the bin size ($\Delta$) of the initial 2D histogram. For our work, we chose $\Delta=0.5$ $\kms$ pixel$^{-1}$. 
Structures of a certain diameter $D$ produce the highest wavelet coefficient at a scale $j$ such that $\Delta 2^j \leq D \leq \Delta 2^{j+1}$. However, they also produce positive (negative) coefficients at other scales, especially if the overdensity (underdensity) is very significant or isolated. 
Other authors focused on the upper scales (e.g., \citealt{Antoja2012,Kushniruk2017}), 
motivated not only by the typical sizes of the structures they were looking for, but also because of limitations imposed by the measurement uncertainties. Now the quality of the data allows the study of smaller scales and thus for this work we investigated the scales $j$=2, 3, 4 and 5, corresponding roughly to the velocity ranges of 2-4 $\kms$, 4-8 $\kms$, 8-16 $\kms$ and 16-32 $\kms$, respectively.


Once the WT was computed, we performed a 
search for local maxima at each scale. 
To that end, we used the method {\tt peak\_local\_max} from the \textit{Python} package {\tt Scikit-image} \citep{Scikit-image}, which returns a list of peaks separated at least \textit{d} pixels. To filter fluctuations that would correspond to a smaller scale, we used a minimum distance $d=2^{j}$ pixels to retain only structures of approximated diameter between $2^{j}$ and $2^{j+1}$. Note that with this approach elongated features will manifest as a trail of peaks, being their precise location along the line mostly determined by fluctuations.

Then, to asses the significance of a peak, we acknowledge that the source of uncertainty in our data (the histogram) is Poisson noise. As in previous work (e.g. \citealt{Antoja2008,Antoja2012}), we selected in the software the option of "Poisson with few events", which is based on the autoconvolution histogram method \citep{Slezak1993} and which enables the study of the low density regions in the velocity plane without loosing precision at 
the centre of the distribution. 
In short, the method evaluates the probability ($P_p$) that a given coefficient is not due to Poisson noise. Since it works in the wavelet space, it takes into account not just the number of stars at the peak, but also in a vicinity whose size, again, depends on the scale.  
Once we specify a significance threshold, the method returns a binary value signifying whether that peak is real or not, at that level of confidence (CL). For this work, we use 4 CL coded as:

\begin{itemize}
\item 0: $P_p < \epsilon_{1\sigma}$
\item 1: $\epsilon_{1\sigma} \leq P_p < \epsilon_{2\sigma}$
\item 2: $\epsilon_{2\sigma} \leq P_p < \epsilon_{3\sigma}$	
\item 3: $P_p\geq \epsilon_{3\sigma}$
\end{itemize} 
where $\epsilon_{n-\sigma}$ corresponds to the area $P[N(0,1) \geq n]$, being N(0,1) the Normal distribution, for instance, $\epsilon_{1\sigma} \approx 0.841$.

Additionally, we coded an alternative metric for the significance of the peaks that is the percentage of times the peak appears in a Bootstrap (BS) of the data. We produced N=1,000 BS of the sample, computed the WT and searched for all the peaks. Afterwards, and for each scale, we counted how many peaks have fallen inside a circle of radius $2^j$ pixels around the peak detected in the data\footnote{Here, we assume that all local maxima correspond to structures of the upper size limit.}. Then, the ratio between this number and the number of BS $N$, gives a value between 0 (the re-sampling failed to reproduce the peak every time) and 1 (the local maximum is always reproduced). 
We refer to this quantity as $P_{BS}$ and is a measure of the probability we had of observing a particular peak in the data.

The output of the method described is a list of local maxima detected at each scale, along with the value of the WT coefficient, 
and the two measures of significance. Then, we consider a peak significant either if $CL\geq2$ or $P_{BS}\geq0.8$. In addition, we counted the number of stars closer to the peak than the minimum distance $d=2^j$ pixels, since those are the source of the positive (negative) wavelet coefficient. For these stars, we also computed their median error in $V_R$ and $V_{\phi}$ to quantify the data uncertainty.


\begin{figure*}[h]
    \centering
    \includegraphics[width=\linewidth]{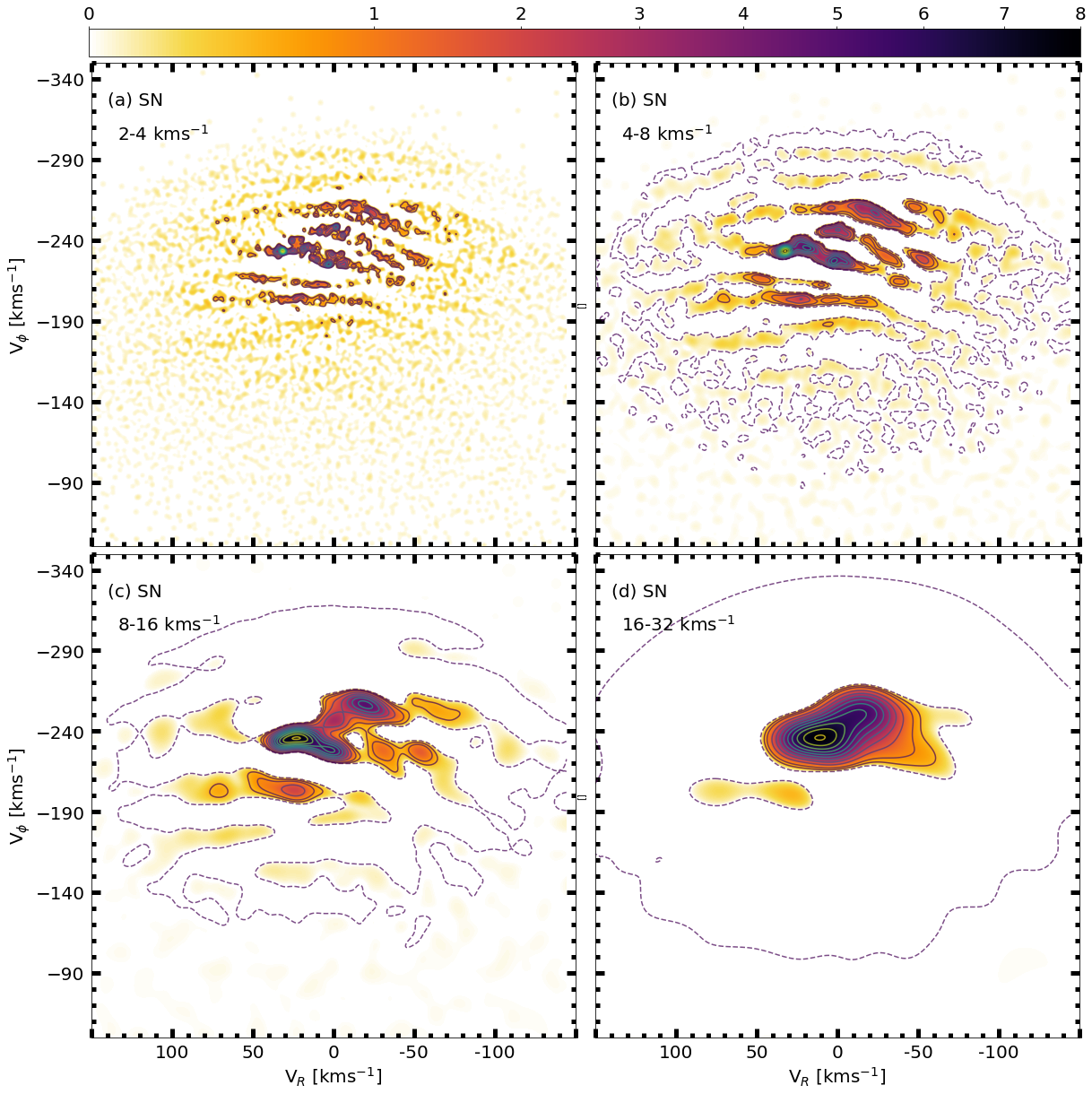}
    \caption{Wavelet planes highlighting the velocity sub-structure in the SN at different scales. The panels correspond to scales from $j=2$ (a) to $j=5$ (d). The colour bar shows only the positive coefficients, just like the solid-line contours, which are shown for different percentages of the maximum coefficient at levels from 10\% to 90\% every 10\%, plus 5\% and 99\%. Additionally, the negative coefficients are represented by the dashed-line contour at the -1\% level. With the values used (see text), the Sun would be located at (-11,-252) $\kms$.}
    \label{fig:SN_wplane}
\end{figure*}

\section{Solar Neighbourhood}
\label{sec:solar}

In this section, we study the kinematic sub-structure of the Solar Neighbourhood sample described in Sect. \ref{sec:sample}. The starting point is the 2D histogram of the Cylindrical velocity coordinates $V_R$ and $V_\phi$, which is very similar to Fig.22 from \cite{Katz2018a}. After applying the methodology presented in Sect.\ref{sec:wavelets} we obtain the wavelet planes (Fig. \ref{fig:SN_wplane}) and a list of significant peaks at the scales of $j=2,3,4,5$, corresponding to 2-4, 4-8, 8-16 and 16-32 $\kms$, respectively. The colours in Fig. \ref{fig:SN_wplane} correspond to the positive WT coefficients, while the contours are presented in two styles: solid lines delimit the positive coefficients at different levels (see caption)  and dashed lines mark the negative coefficients at a single level. The negative contours help track the valleys and highlight the elongated features, but we did not plot them for the scale $j$=2 to ease the visualization.

Panels (c) and (d) in Fig. \ref{fig:SN_wplane} show a bi-modality and an asymmetric yet structured central distribution formed by several large-scale kinematic groups, similarly to previous studies \citep[e.g.][]{Dehnen1998,Antoja2012}. However, at lower scales (panels (a) and (b)) we see the recently discovered thin arches \citep{Katz2018a} crossing the velocity space in a nearly horizontal direction. Next we analyse these thin arches (Section \ref{subsec:arches}) and the larger scale kinematic groups (Section \ref{subsec:groups}) in more detail.


\subsection{Thin arches in the velocity distribution}
\label{subsec:arches}
\begin{figure*}
    \centering
    \includegraphics[width=\linewidth]{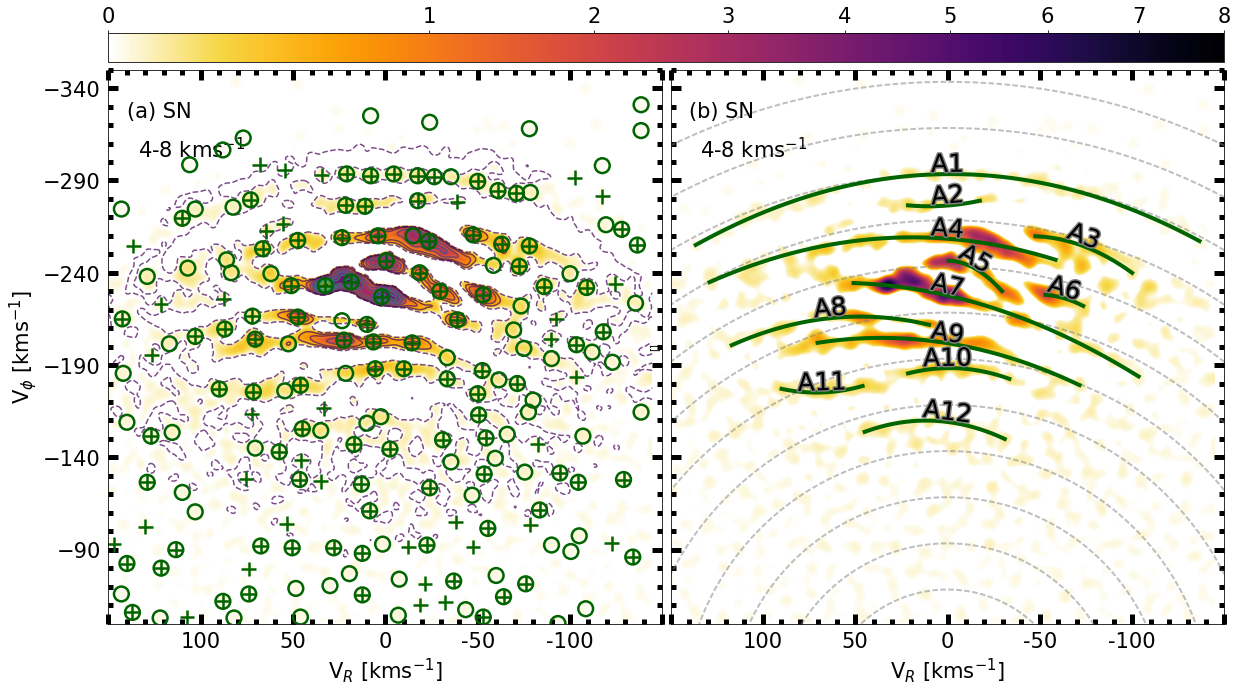}
    \caption{Wavelet plane at small scale ($j$=3) with the peaks and arches found. The circles correspond to those local maxima with a Poisson significance greater or equal to 2 ($P_p \geq \epsilon_{2\sigma}$), whereas crosses correspond to those with $P_{BS} \geq 0.8$. In both cases, the diameter corresponds to $\Delta 2^{j+1}$, indicating the highest size expected for the structures at this scale. On the right, we have associated the peaks into arches and fitted a parabola (dark green lines). The dashed grey lines then correspond to constant kinetic energy tracks.}
    \label{fig:SN_j3}
\end{figure*}

\begin{table*}
\caption{Arches found at small scale $j$=3 (4-8 $\kms$). The arches are those appearing in Fig.\,\ref{fig:SN_j3}, ordered by increasing $V_\phi$. For each arch we give a label in the first column, the Classical MGs that it contains (if any) in the second column, and 3 of its points in the velocity plane in columns 3, 4 and 5 respectively, the coordinates of the first and last peak of the line, and the expected azimuthal velocity corresponding to zero radial velocity. The last column shows a reduced $\chi^2$ between the arch and the corresponding track of constant kinetic energy.}
\label{tab:scale3}
\centering
\begin{tabular}{llcccc}
\hline\hline
n. & Contains & Start (V$_R$,V$_\phi$)& End (V$_R$,V$_\phi$) & 0-cross & Reduced $\chi^2$\\
\hline
A1 & - & 136.5, -254.5 & -136.5, -255.0 & -293.4 & 4.09 \\
A2 & - & 21.5, -276.5 & -17.5, -279.0 & -276.4 & 1.20 \\
A3 & $\gamma$Leo& -47.5, -260.5 & -100.0, -239.5 & -246.5 & 335.05 \\
A4 & Sirius& 129.0, -238.0 & -58.5, -244.0 & -258.4 & 25.52 \\
A5 & Coma,Dehnen98-6& -0.5, -246.5 & -29.5, -230.0 & -246.5 & 44.96 \\
A6 & Dehnen98-14& -53.0, -228.0 & -73.5, -222.0 & -200.4 & 1290.04 \\
A7 & Hyades,Pleiades& 51.0, -233.0 & -103.5, -183.5 & -227.7 & 111.01 \\
A8 & - & 117.0, -201.5 & 10.0, -212.0 & -209.1 & 297.48 \\
A9 & Hercules, $\epsilon$Ind& 70.5, -204.0 & -71.5, -180.0 & -201.9 & 42.84 \\
A10 & HR1614& 21.5, -185.5 & -33.5, -182.5 & -188.3 & 1.14 \\
A11 & Arifyanto05& 90.0, -177.0 & 46.0, -179.0 & -203.7 & 260.49 \\
A12 & Arcturus& 45.0, -155.5 & -31.0, -149.5 & -159.2 & 5.50 \\
\hline
\hline
\end{tabular}
\end{table*}

The precision of the \textit{Gaia} data, with uncertainties in the measurements well below 1 $\kms$, and the large number of stars in the SN sample (Table \ref{tab:sectors}) allows us for the first time to  appreciate unprecedented details of the velocity plane at small scales. The existence of thin arches in the \textit{Gaia} data was first discovered by simple inspection of two-dimensional histograms of the velocity distribution. Here, the WT gives us a new representation of the data, allowing for a characterization of the arches and, more importantly, the evaluation of their statistical significance. 

Panels (a) and (b) in Fig. \ref{fig:SN_wplane} clearly show wavelet coefficients that appear connected and organized in the mentioned thin arches, either composed of overdensities (positive coefficients) or underdensities (negative coefficients). The scales at which these elongated features appear indicate that they are intrinsically thin, at the level of 2-8 $\kms$. Another advantage of the WT is that it highlights overdensities otherwise hard to discern by direct inspection of the histogram, like the arches at high |$V_\phi$|. Although panels (a) and (b) unveil similar kinematic sub-structure, 
some of the arches appear to be split into two at the lower scale, for instance, the structure passing through ($V_R$, $V_\phi$)=(0,-270) $\kms$. However, for the sake of simplicity, the rest of this sub-section is devoted to the scale $j$=3 alone.

Figure \ref{fig:SN_j3}a shows the peaks detected at scale of 4-8 $\kms$. Their coordinates and characteristics are available in the on-line data material (see Appendix \ref{App:material}). The ones signalled with a circle are significant in terms of the Poisson noise (significance greater or equal to 2, $P_p \geq \epsilon_{2\sigma}$). Peaks marked with a cross are those that appeared in more than 80\% of the Bootstraps. As already discussed in Sect. \ref{sec:wavelets}, in the case of elongated structures our peak finder procedure forcedly yields a trail of peaks. Although this method is suboptimal in detecting entire arches, we can nonetheless indirectly detect them as well as define their shape in the following manner. We selected as members of each arch the significant peaks, according either to their CL or their $P_{BS}$, that show continuity in the WT coefficient. With this definition, an arch ends when the significance of the peaks we detect drops below the threshold. 
We note that in some cases where the connectivity of the wavelet coefficients is not clear, this grouping is slightly conjectural, yet our global conclusions do not vary. Once the peaks were selected and grouped, we fitted a second order polynomial, $\tilde{V}_\phi(V_R)$, to each arch between its minimum and maximum $V_R$. 

The arches obtained are plotted in Fig. \ref{fig:SN_j3}b with solid dark green lines and their characteristics are given in Table \ref{tab:scale3}, including the coordinates in velocity space of the first and last peaks, as well as the zero-order term of the polynomial (which corresponds to $\tilde{V}_\phi(V_R=0)$). In the table we also include, for reference, the name of previously known kinematic groups (discussed in Sect. \ref{subsec:groups}) that are contained within the arches. However, it is important to remark that, since past studies dealt with more roundish moving groups (due mostly to the lack of precision in the astrometry), the Classical moving groups (MG) and our arches are morphologically different entities. 

We detected 12 arches, 6 of which cover a large range of radial velocities (A1,4,7,8,9,12). All of them are roughly constant in azimuthal velocity. We also see that, while some parabolas are symmetrical with respect to $V_R=0$, others are inclined and/or present the maximum at $V_R\neq0$, as already noted by \cite{Katz2018a}. We confirm the existence of A1 already reported by \cite{Katz2018a} while A8 and A9 would correspond to the two branches of Hercules described in the same study. 
From all the underdensities seen in this velocity space, the most prominent known one (the gap separating Hercules from the central part) is now clearly seen running between A7 and A8-9. Another gap, just below A12, hints to a 13th arch although the small number of peaks in panel (b) do not allow us to constrain it. In contrast, we see a well organized trail of peaks above A1, at $V_\phi \sim -320$ $\kms$, but these only have 1 star each and therefore we refrain from connecting them as an arch.

Several models have shown that dynamical processes like phase mixing can imprint "arc-like features" of roughly constant kinetic energy in the velocity plane \citep{Minchev2009,Gomez2012a}, and because of this we included constant energy profiles in Fig. \ref{fig:SN_j3}b as grey dashed lines. We see how some of the arches found, e.g. A1, are clearly compatible with having a constant kinetic energy along its trail. To quantify this similarity, we compared each parabola (evaluated with a step size of $0.1$ $\kms$) to the corresponding constant kinetic energy line, using the energy of the parabola at zero radial velocity, i.e. $\frac{1}{2}\tilde{V}^2_\phi(V_R=0)$., and calculated a reduced $\chi^2$ statistic (last column in Table \ref{tab:scale3}). For the short arches, based on the few peaks used for their fit, this quantity may not be representative. 
For the rest, according to the reduced $\chi^2$ arches 1, 10 and 12 are the most compatible, while A3, 7 and 8 show the largest discrepancy with a constant energy line.


\subsection{Classic and new kinematic groups}
\label{subsec:groups}

\begin{figure}
\centering
\includegraphics[width=\linewidth]{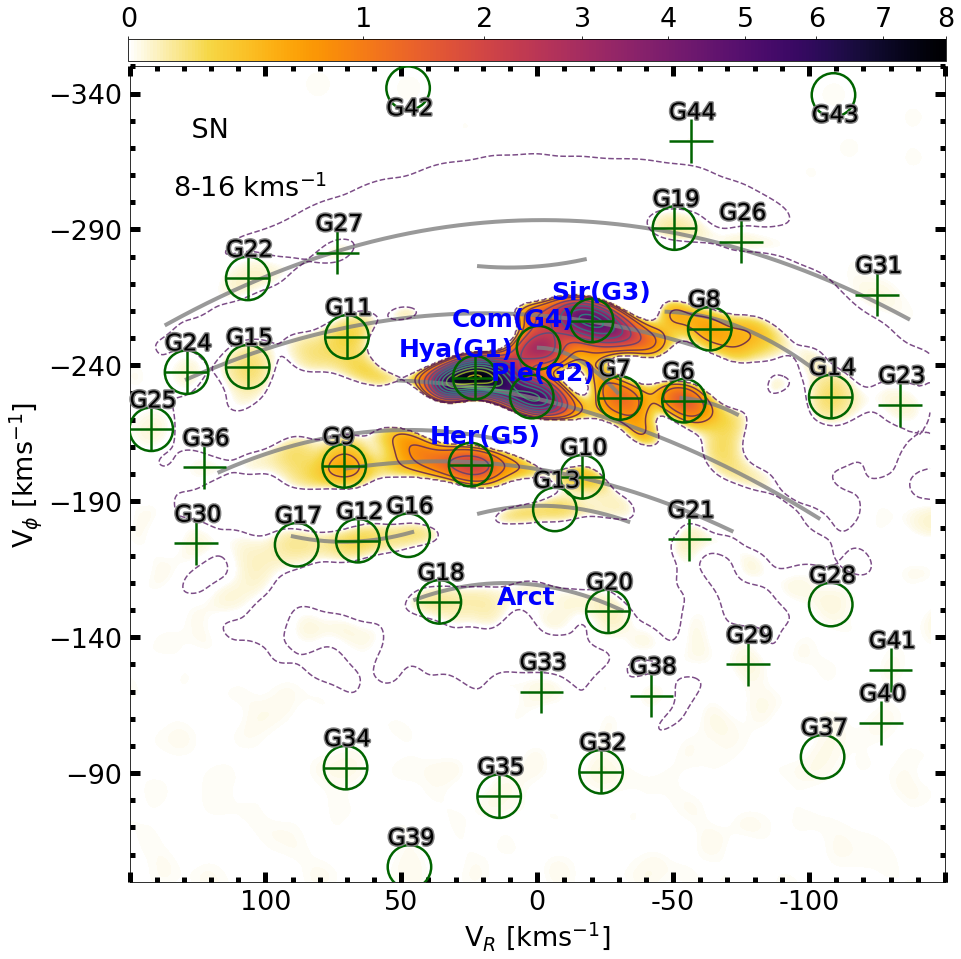}
\caption{Wavelet plane ($j$=4) with the peaks found. The circles and crosses are defined as in Fig. \ref{fig:SN_j3}. Also, solid grey lines represent the arches from the same figure.}
    \label{fig:SNj4_peak}
\end{figure}

\begin{table*}

\caption{Groups detected at scale $j$=4 ($\sim$8-16 $\kms$). We ordered the objects by decreasing order of Wavelet coefficient (column 5), as indicated by the index in column 1. For each object, we show the velocity coordinates of the peak (columns 3 and 4), as well as the two significance statistics (see Sect. \ref{sec:wavelets}) in columns 6 and 7. In addition, columns 8 to 10 contain the number of stars inside a circle of diameter 16 $\kms$ along with the associated median error. The last column contains the number of known structures that matched the coordinates of the peak, from which we select the one that appears the most as 'Name' in column 2 (see Appendix \ref{App:MG} for details). We found 4 cases of multiple-matching. The Pleiades have another match which is, in fact, a MG from \cite{Xia2015} called "Hyades-Pleiades". In the case of Hercules II, the other source is Dehnen98-8. Something similar happens with $\gamma$Leo, for the second match is Dehnen98-13. Finally, Dehnen98-6 overlaps with Wolf 630.}
\label{tab:scale4}
\centering

\begin{tabular}{llrccccccccc}
\hline\hline
 n. &            Name &  $V_R$ &  $V_\phi$ &  Wavelet &  CL &  $P_{BS}$ &  Stars &  Median $\sigma_{V_R}$ &  Median $\sigma_{V\phi}$ &  Matches \\
\hline
  G1 &          Hyades &   23.0 &    -235.5 &   7.6575 &     3 &      1.00 &  22422 &                   0.4 &                      0.5 &        1 \\
  G2 &        Pleiades &    2.0 &    -228.5 &   5.6147 &     3 &      0.58 &  21524 &                   0.3 &                      0.5 &        2 \\
  G3 &          Sirius &  -20.0 &    -256.5 &   4.9261 &     3 &      1.00 &  17285 &                   0.3 &                      0.5 &        1 \\
  G4 &  Coma Berenices &   -0.5 &    -246.5 &   3.0140 &     3 &      0.72 &  19330 &                   0.3 &                      0.5 &        1 \\
  G5 &     Hercules II &   24.5 &    -203.5 &   2.0481 &     3 &      1.00 &   7602 &                   0.4 &                      0.6 &        2 \\
  G6 &  Dehnen98-14 &  -54.0 &    -227.0 &   1.5017 &     3 &      1.00 &   6331 &                   0.4 &                      0.5 &        1 \\
  G7 &   Dehnen98-6 &  -30.5 &    -228.0 &   1.2290 &     3 &      1.00 &  10463 &                   0.3 &                      0.5 &        2 \\
  G8 &        $\gamma$Leo &  -63.5 &    -253.5 &   0.5571 &     3 &      0.96 &   3359 &                   0.5 &                      0.4 &        2 \\
  G9 &         $\epsilon$Ind &   71.0 &    -203.0 &   0.5357 &     3 &      1.00 &   2192 &                   0.6 &                      0.6 &        1 \\
 G10 &     Liang17-9 &  -16.5 &    -199.0 &   0.2773 &     3 &      1.00 &   3805 &                   0.2 &                      0.6 &        1 \\
 G11 &    Antoja12-GCSIII-13 &   70.0 &    -250.5 &   0.1813 &     3 &      0.84 &   1392 &                   0.7 &                      0.4 &        1 \\
 G12 &            GMG 1 &   66.0 &    -175.5 &   0.1510 &     3 &      0.98 &    900 &                   0.7 &                      0.7 &        0 \\
 G13 &            GMG 2 &   -6.5 &    -187.0 &   0.1266 &     3 &      0.52 &   2672 &                   0.2 &                      0.7 &        0 \\
 G14 &   Antoja12-12 & -108.0 &    -228.5 &   0.1188 &     3 &      1.00 &    526 &                   0.7 &                      0.5 &        1 \\
 G15 &   Antoja12-16 &  106.5 &    -239.5 &   0.1119 &     3 &      1.00 &    376 &                   0.9 &                      0.5 &        1 \\
 G16 &     Arifyanto05 &   47.5 &    -177.5 &   0.0928 &     3 &      0.66 &   1260 &                   0.5 &                      0.7 &        1 \\
 G17 &            GMG 3 &   88.5 &    -174.0 &   0.0699 &     3 &      0.77 &    428 &                   0.7 &                      0.7 &        0 \\
 G18 &         $\eta$Cep (Arct)&   36.0 &    -153.0 &   0.0596 &     3 &      0.89 &    445 &                   0.5 &                      0.8 &        1 \\
 G19 &            GMG 4 &  -50.5 &    -290.5 &   0.0553 &     2 &      1.00 &    270 &                   0.5 &                      0.6 &        0 \\
 G20 &   Antoja12 (17) (Arct)&  -26.0 &    -149.5 &   0.0305 &     2 &      0.88 &    291 &                   0.3 &                      0.9 &        1 \\
 G21 &            GMG 5 &  -56.0 &    -176.0 &   0.0275 &     1 &      0.92 &    452 &                   0.5 &                      0.8 &        0 \\
 G22 &            GMG 6 &  106.5 &    -272.0 &   0.0248 &     3 &      0.99 &     69 &                   1.1 &                      0.6 &        0 \\
 G23 &            GMG 7 & -133.5 &    -225.5 &   0.0164 &     1 &      0.94 &     86 &                   0.9 &                      0.5 &        0 \\
 G24 &            GMG 8 &  129.0 &    -237.5 &   0.0155 &     2 &      0.98 &     72 &                   1.0 &                      0.5 &        0 \\
 G25 &            GMG 9 &  142.0 &    -216.5 &   0.0131 &     2 &      0.96 &     46 &                   1.4 &                      0.6 &        0 \\
 G26 &           GMG 10 &  -75.0 &    -285.5 &   0.0120 &     0 &      0.83 &    109 &                   0.6 &                      0.5 &        0 \\
 G27 &           GMG 11 &   73.5 &    -281.5 &   0.0110 &     0 &      0.88 &    145 &                   0.7 &                      0.5 &        0 \\
 G28 &           GMG 12 & -108.0 &    -152.0 &   0.0107 &     2 &      0.76 &     41 &                   0.8 &                      1.0 &        0 \\
 G29 &           GMG 13 &  -77.5 &    -130.0 &   0.0101 &     1 &      0.81 &     52 &                   0.6 &                      1.0 &        0 \\
 G30 &           GMG 14 &  125.5 &    -174.5 &   0.0100 &     1 &      0.86 &     74 &                   1.0 &                      0.7 &        0 \\
 G31 &           GMG 15 & -125.0 &    -266.0 &   0.0092 &     1 &      0.97 &     23 &                   1.0 &                      0.4 &        0 \\
 G32 &           GMG 16 &  -23.5 &     -90.5 &   0.0084 &     2 &      0.91 &     24 &                   0.5 &                      1.4 &        0 \\
 G33 &           GMG 17 &   -1.5 &    -120.0 &   0.0082 &     1 &      0.84 &     63 &                   0.4 &                      1.1 &        0 \\
 G34 &           GMG 18 &   70.5 &     -92.0 &   0.0067 &     2 &      0.85 &     19 &                   0.9 &                      1.9 &        0 \\
 G35 &           GMG 19 &   14.0 &     -81.5 &   0.0063 &     2 &      0.86 &     20 &                   0.4 &                      1.1 &        0 \\
 G36 &           GMG 20 &  122.5 &    -202.5 &   0.0060 &     0 &      0.86 &     94 &                   0.9 &                      0.5 &        0 \\
 G37 &           GMG 21 & -105.0 &     -96.0 &   0.0057 &     2 &      0.72 &     15 &                   1.1 &                      1.6 &        0 \\
 G38 &           GMG 22 &  -42.0 &    -118.5 &   0.0055 &     0 &      0.89 &     53 &                   0.4 &                      1.2 &        0 \\
 G39 &           GMG 23 &   47.0 &     -55.5 &   0.0046 &     2 &      0.61 &     11 &                   0.6 &                      1.1 &        0 \\
 G40 &           GMG 24 & -126.5 &    -108.5 &   0.0040 &     1 &      0.81 &     13 &                   1.0 &                      0.9 &        0 \\
 G41 &           GMG 25 & -130.0 &    -128.0 &   0.0040 &     1 &      0.80 &      8 &                   0.6 &                      0.9 &        0 \\
 G42 &           GMG 26 &   47.5 &    -342.0 &   0.0013 &     2 &      0.64 &      1 &                   2.0 &                      2.2 &        0 \\
 G43 &           GMG 27 & -109.0 &    -339.5 &   0.0013 &     2 &      0.57 &      1 &                   1.0 &                      1.1 &        0 \\
 G44 &           GMG 28 &  -56.5 &    -322.5 &   0.0009 &     0 &      0.84 &      2 &                   1.0 &                      1.1 &        0 \\
\hline
\hline
\end{tabular}

\end{table*}

From the wavelet planes shown in panels (c) and (d) of Fig. \ref{fig:SN_wplane}, we can see the dominance of more rounded structures as the elongated features merge or fade away. Nonetheless, in panel (c) we can still appreciate some imprints of elongated structures. From both panels, we can clearly recognize a bi-modality formed by the centre of the distribution and the Hercules moving group. Yet, in panel (d) the latter presents a peanut-shape, which is in fact related to two different structures, while the central over-density contains the Classical MGs.

In Fig. \ref{fig:SNj4_peak} we show the significant peaks found at $j$=4 where, just like before, circles are related to Poisson and crosses, to Bootstraps. For reference, we also plot the arches from  Fig. \ref{fig:SN_j3}b as grey lines. It becomes obvious that most of the peaks sit on top of an arch. 
It is to be expected, then, for some of the known MGs to belong to the same arch, as we shall see below. All the detections in Fig. \ref{fig:SNj4_peak} are listed in Table \ref{tab:scale4} ranked according to their wavelet coefficient, i.e. their relative strength at that scale. The list contains also the coordinates of the peak and the two measures of significance. The names of the objects are given according to the name in the literature that is most commonly assigned to the group (see Appendix \ref{App:MG} for details). Finally, if no match is found, we assign a new name composed of the letters 'GMG' (\textit{Gaia} Moving Group) and a number according to their position on the list. The table also contains the number of stars contributing to the peak, along with their median errors in the radial and azimuthal velocities.

\begin{figure*}
    \centering
        \includegraphics[width=0.98\textwidth]{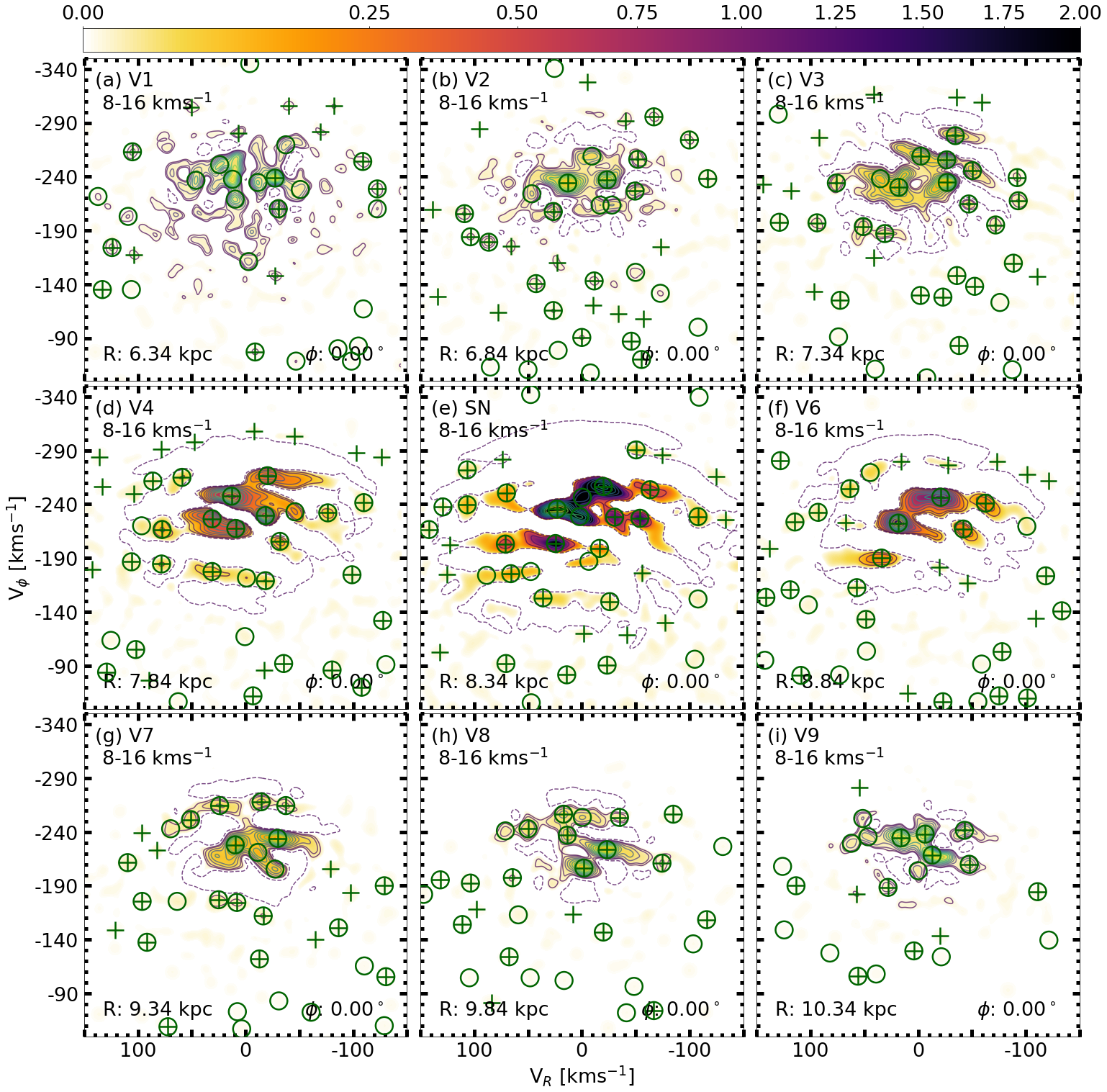} 

\caption{Wavelet planes of different Galactic neighbourhoods along the zero-azimuth line. Panels (a) to (i) correspond to the volumes V1 to V9 described in Table \ref{tab:sectors}. We show the coefficients in a common colour bar with the significant peaks superposed and contours in the same manner as in Fig. \ref{fig:SN_wplane}. We notice the change in colour caused by the difference in number of stars of each sub-sample. Also, for the SN sample, the most crowded regions of the plane saturate (c.f. Fig.\ref{fig:SNj4_peak}).}
\label{fig:1exp_R}
\end{figure*}

As can be seen, the known moving groups are among the peaks with higher coefficients, consistent with previous samples. 
The most noticeable differences are the wavelet coefficients and peaks at the less dense regions on the velocity plane, which {\it Gaia} samples now with significantly better statistics. For instance, the Arifyanto05 (G16) appears to be elongated and matches one of the found arches (A11), plotted in grey beneath. Its relation with groups II and III from \cite{Helmi2006}, however, is still not clear. Also, Arcturus (associated to G18 and G20) is very conspicuous, while before was only detected in especially selected samples \citep{Williams2009}. The larger wavelet coefficients for G18 compared to G20, are consistent with previous findings that Arcturus has a preferential positive $V_R$ (negative $U$). Besides, $\epsilon$Ind (G9) appears to be more connected to Hercules than in previous studies \citep{Liang2017}, and similarly for Liang17-9 (G10). Regarding Hercules, in contrast with other studies of the Solar Neighbourhood, we detect a single peak between the two reported previously \citep[e.g.,][]{Antoja2012,Bobylev2016}. 
Also interesting is the connection between  $\gamma$Leo (G8) and Antoja12-12 (G14), which was not observed in \cite{Antoja2012}, following one of the arches (A3). 

The most noticeable change with respect to past work, however, is that $\sim$30 candidates to moving groups are found, roughly half of which have a CL$\geq$2. A more exhaustive cross-match with the literature is required to find previously detected groups among them and their definitive existence needs to be confirmed since the significance is not high enough. However, four of the new groups have CL$\geq$3 (GMG1-3 and GMG6), and, remarkably, we find four new groups (GMG-1 to -4) with strength comparable to that of the known MGs. While some of them may very well be part of the already recognized kinematic structures, (e.g. G12 (GMG1) and G17 (GMG3) related to Arifyanto05, and  G13 (GMG2) related to Liang2017-9), peak G19 (GMG4) is part of the newly characterized arch (A1).  


\section{Unexplored Galactic neighbourhoods}
\label{sec:vicinity}

\begin{figure*}
    \centering
        \includegraphics[width=0.98\textwidth]{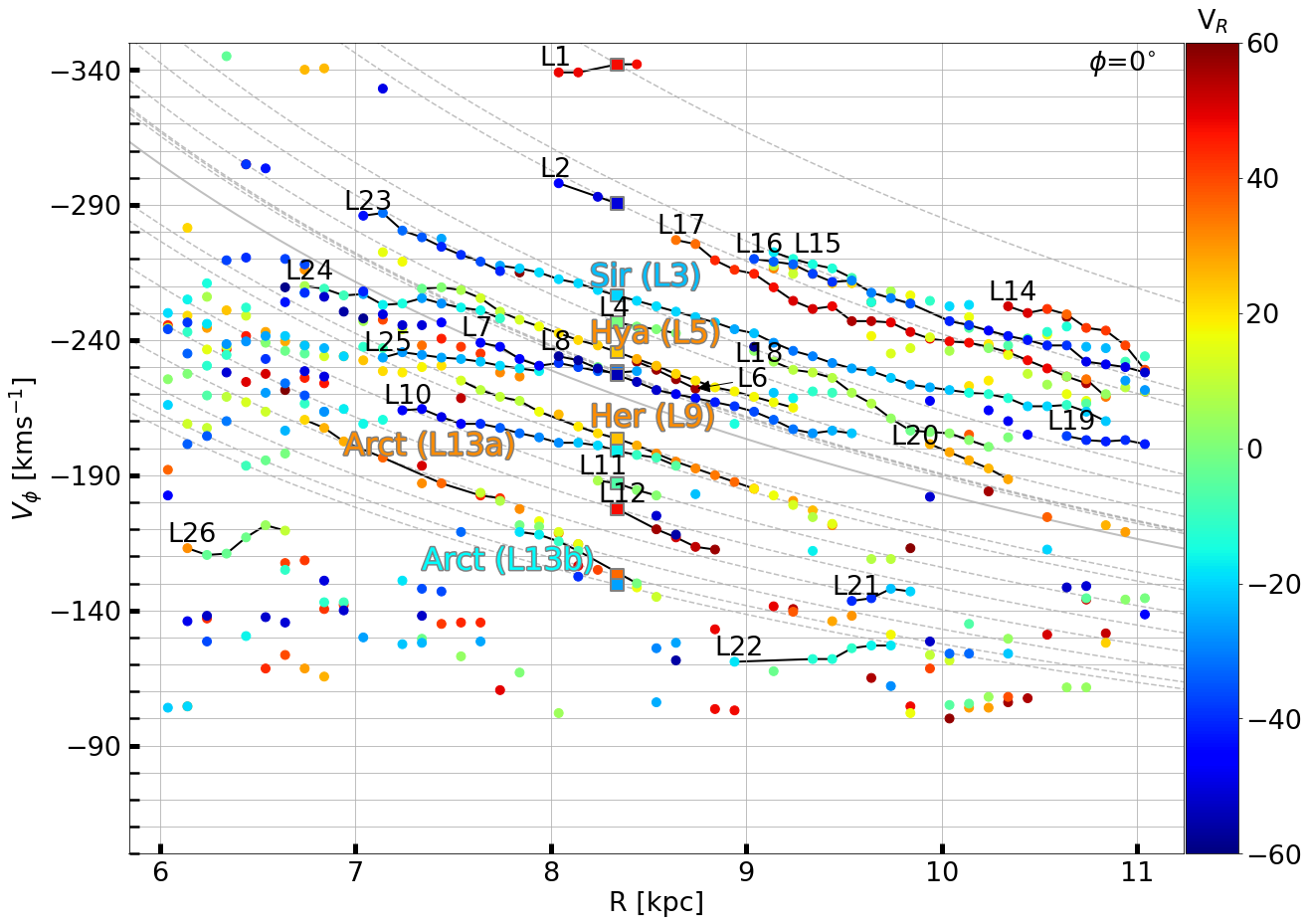} 

\caption{Evolution of the rotation velocity of the kinematic sub-structures as a function of the radius. From all the peaks at the 51 volumes along the Sun-GC line, only significant peaks according to Poisson (CL$\geq$2) with radial velocities between [+60,-60] $\kms$ have been considered. The peaks detected at Solar radius (8.34 kpc) are shown as squares. The rest, are shown as circles, linked together through solid black lines when observed to belong to the same structure (except for L13, see text). The dashed grey lines correspond to constant angular momentum tracks.
}
\label{fig:2exp_R}
\end{figure*}

\begin{table*}
\caption{Lines marked in Fig. \ref{fig:2exp_R}. The first column contains the label of the Line, while the second refers to the name of the structure at Solar radius (if possible). Then, columns 3 and 4 contain the Arches (Table \ref{tab:scale3}) and Groups (Table \ref{tab:scale4}) corresponding to the peak at Solar radius of the Line. The last three columns characterize the lines by, respectively, their number of points, the slope (kms$^{-1}$kpc$^{-1}$) obtained from the linear regression and the corresponding correlation coefficient, $r$.}
\label{tab:linesR_Vphi} 
\centering
\begin{tabular}{llllccc}
\hline\hline
n. & Name & Arch & Group & Points & Slope & r  \\
\hline
L1 &  & & G42& 4 & -9.0$\pm$2.1& -0.95 \\
L2 & Arch 1& A1& G19& 3 & 25.0$\pm$0.0& 1.00 \\
L3 & Sirius& A4& G3& 32 & 19.2$\pm$0.4& 0.99 \\
L4 & Coma Berenices& A5& G4& 4 & 13.0$\pm$0.7& 1.00 \\
L5 & Hyades& A7& G1& 20 & 24.9$\pm$0.5& 1.00 \\
L6 & & & & 4 & 33.5$\pm$0.9& 1.00 \\
L7 & Dehnen98-6& A5& G7& 9 & 13.3$\pm$2.2& 0.92 \\
L8 & Dehnen98-14& A6& G6& 16 & 20.0$\pm$0.7& 0.99 \\
L9 & Hercules II& A9& G5& 15 & 26.5$\pm$0.2& 1.00 \\
L10 & Liang17-9& A9& G10& 15 & 14.3$\pm$0.5& 0.99 \\
L11 & HR1614 & A10 & G13& 4 & 19.0$\pm$2.1& 0.99 \\
L12 & Arifyanto05 & A11 & G16 & 5 & 31.0$\pm$2.7& 0.99 \\
L13a & Arcturus& A12& & 7 & 29.8$\pm$1.5& 0.99 \\
L13b & Arcturus& A12& & 5 & 32.9$\pm$3.4& 0.98 \\
L14 & & & & 8 & 30.2$\pm$5.2& 0.92 \\
L15 & & & & 5 & 22.5$\pm$1.7& 0.99 \\
L16 & & & & 20 & 22.0$\pm$0.4& 1.00 \\
L17 & & & & 24 & 23.4$\pm$0.8& 0.99 \\
L18 & & & & 12 & 31.0$\pm$1.8& 0.98 \\
L19 & & & & 5 & 6.0$\pm$1.9& 0.88 \\
L20 & & & & 5 & 32.0$\pm$1.2& 1.00 \\
L21 & & & & 4 & -14.0$\pm$5.9& -0.86 \\
L22 & & & & 6 & -8.5$\pm$2.5& -0.86 \\
L23 & & & & 8 & 31.4$\pm$2.0& 0.99 \\
L24 & & & & 11 & 10.0$\pm$1.3& 0.93 \\
L25 & & & & 8 & 7.9$\pm$1.2&0.92 \\
L26 & & & & 6 & -20.4$\pm$6.8& -0.83 \\
\hline
\hline
\end{tabular}
\end{table*}

In this section, we focus on the abundant kinematic sub-structure that we find in different Galactic neighbourhoods and its variation with azimuth and radius. 
For this exploration we used the 216 sub-samples described in Sect. \ref{sec:sample} (golden dots in Fig.\ref{fig:sample}). 
Most of our exploration in this section is based on the scale of 8-16 $\kms$ since smaller scales look more noisy at the distant neighbourhoods for several reasons: the number of stars decreases as we move away from the Sun, the velocity uncertainties increase and the velocity dispersion at inner Galactic radius is larger, effectively decreasing the density of stars in the velocity plane.

Figure \ref{fig:1exp_R} shows the wavelet plane at scale of 8-16 $\kms$ for a small sample of different Galactic neighbourhoods located at the the Solar azimuth ($\phi=0\deg$) but at different radius going from 6.34 (V1) to 10.34 kpc (V9). It is not necessary to move very far, though, to notice changes in the sub-structure of the velocity plane.
For instance Hercules, located at $V_\phi\sim200$ $\kms$ in the SN, appears to be shifted to -220 $\kms$ in V4 and to -190 $\kms$ in V6. 
 In fact, we can trace this structure from V4 to V7, but in V3 it seems to have mixed with the central part of the velocity distribution. This shift is consistent with previous findings \citep[e.g.][]{Antoja2012} but now we trace it farther from the Sun than ever before. The same behaviour, shifting to smaller $|V_\phi|$ when increasing $R$, is also observed for other structures like Sirius and Hyades. In addition, a prominent arched structure at $V_\phi\sim-260$ $\kms$ and positive $V_R$ is observed at V6, then more strongly at the V7 and V8 volumes. At the SN, though, it appears quite weak and only represented by the two peaks that trace A1 from Fig. \ref{fig:SN_j3}. 
 
The changes of the velocity sub-structure and specially its smooth evolution with radius can be better seen in an animation (on-line animation 1, see Appendix \ref{App:material}) which considers now the 51 neighbourhoods at different radius with at a step of 100 pc and extending in $R$ from $\sim$6 kpc to $\sim$ 11 kpc. In this animation, the downwards movement of the different kinematic structures when increasing distance from the Galactic centre can be perfectly followed as continuous passages of kinematic waves. The equivalent animations at the scales of 4-8 (on-line animation 2) and 16-32 $\kms$ (on-line animation 3) show a similar evolution. 

To quantify the changes in the velocity distribution with Galactocentric radius, we collected all peaks detected in the WT at scale $j=4$ with a CL$\geq$2 in the 51 different volumes presented in Sect. \ref{sec:sample}, and plotted in Fig.\ref{fig:2exp_R} their $V_\phi$ as a function of $R$ and colour-coding the symbols according to radial velocity. With the help of the colour code, we associated all peaks that seemed to follow a common trend (both in $V_\phi$ and $V_R$) and linked them with a continuous solid black line, each labelled with a number. While this association of peaks may seem rather subjective, the continuity in most of the cases is beyond doubt. 
A simple inspection of this figure reveals a wide variety of peaks forming a general trend downwards, consistent with a decreasing $|V_\phi|$ with $R$ as mentioned above. The expected uncertainties in the location of the peaks along the vertical axis are negligible in front of the variations observed\footnote{Even though the algorithm we use does not return an error associated to the location of the peak, this can be approximated by the error on the mean velocity of the $N$ stars composing each group: we take the maximum between their standard deviation and their uncertainty in velocity, and then divide by the square root of $N$. These errors are between 0.04 $\kms$ and 1.06 $\kms$ for 80\% of the peaks in the Solar Neighbourhood and between $\sim$0.3 - 5.0 $\kms$ on average for neighbourhoods at 1 kpc from the Sun, with the low |$V_\phi$| region of the velocity plane being dominated by the larger errors.}. We note that lines span a wide range of Galactocentric radius, meaning that the peaks at the SN have their counterparts in many other Galactic neighbourhoods, but at a different azimuthal velocity. Table \ref{tab:linesR_Vphi} lists all the prominent lines in Fig.\ref{fig:2exp_R} where we also indicate the name of the kinematic arch or group that could be linked to each line, based on the peaks found at Solar radius (marked with squares instead of circles). Moreover, we used a linear regression to estimate the slope of the line, along with its associated uncertainty\footnote{In this calculation, the individual uncertainties of each peak have not been taken into account.} (penultimate column in Table \ref{tab:linesR_Vphi}). Overall, excluding those lines that deviate from the main trend (L1,21,22,26), we observe a mean slope of 23$\pm$2 kms$^{-1}$kpc$^{-1}$. 

By inspecting Table \ref{tab:linesR_Vphi}, we note some scatter in the measured slopes. Along with Fig.\ref{fig:2exp_R}, we see that there seems to be two different behaviours in the slope of the most prominent lines: L5 and L9 have slopes $\sim$25 kms$^{-1}$kpc$^{-1}$, while L3, 10, 24 and 25 have smaller slopes, between 10 and 20 kms$^{-1}$kpc$^{-1}$. Developing on this, since the resonant effects of the bar and spiral arms are expected to create kinematic sub-structure with approximately constant angular momentum $L_Z=RV_\phi$ \citep{Sellwood2010,Quillen2018}, we over-plot in Fig.\ref{fig:2exp_R} lines of constant $L_Z$ (dashed grey lines). As a result, we see how L5 and L9 follow rather well the lines of constant $L_Z$, while other lines such as L3 and L10 do not. 
Moreover, the colour of the peaks reveal variations in radial velocity, which is one of the perquisites of using the WT compared to a direct plot of $V_\phi$ against $R$. For most of the lines, this gradient (or oscillation in some cases) might be related to noise but, as we shall see in the discussion, for some cases like Hercules, this extra information can help us understand the origin of such structure.

\begin{figure*}
    \centering
        \includegraphics[width=0.98\textwidth]{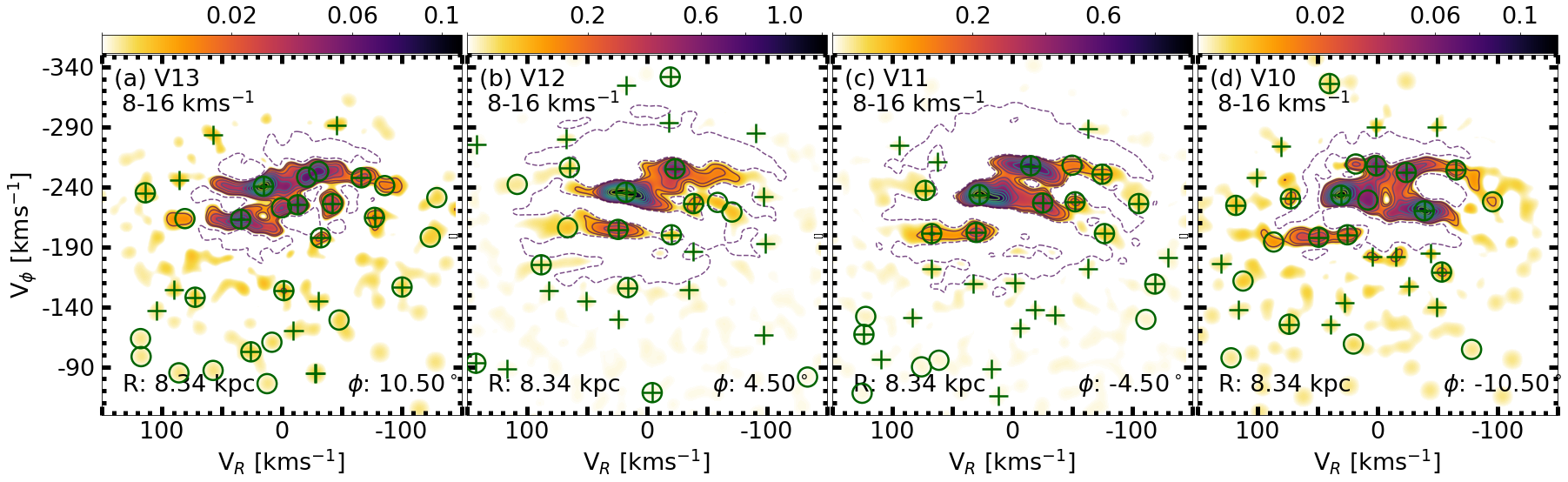} 
\caption{Wavelet planes of different Galactic neighbourhoods along the arc at Solar radius. Panels (a) to (d) correspond to the volumes V13 to V10 described in Table \ref{tab:sectors}. The contour levels and the symbols are defined in the same manner as in Fig. \ref{fig:SN_wplane}.}
\label{fig:1exp_P}
\end{figure*}

Focusing now on individual structures that traverse the Solar radius, L3 (Sirius) is the line with the largest reach, extending from $R\sim7.5$ to $R\sim11$ kpc. Other groups with large extensions are Hyades (L6), Dehnen98-6 and Dehnen98-14 (respectively, L7 and L8) or Hercules (L9). Next to Hercules in the SN, is the moving group Liang17-9 (one of the several peaks linked to A9), represented here by L10. The evolution of this group with radius differs significantly from that of Hercules, suggesting these two sub-structures being different entities and having a different origin. Finally, bellow those two structures, we have Arcturus which, as can be seen, has been split into two lines, L13a and L13b. This is because the elongated nature of this structure, as already mentioned, makes our algorithm  detect the two ends of the structure separately. We notice that, even though we detected Pleiades at SN, we do not observe any continuity with radius for this group since its relation with L25 or L7 is unclear.

We also find clear lines that do not cross the Solar radius, that is, new peaks that are detected only in extrasolar volumes. The clearest examples at outer Galactic neighbourhoods are L15 to 17, which correspond to the prominent arch observed in panels (g) and (h) of Fig. \ref{fig:1exp_R}. Their connection with L2 is only truncated at 2 volumes, yet traceable in Fig. \ref{fig:1exp_R}, and the splitting can be explained using the same argument as for Arcturus. As for L14 and L18, there is no clear link to any of the structures detected at SN and nonetheless, they evolve similarly to the lines between them. There are also prominent new structures at inner radii such as L23 or L24, which could be associated by visual inspection of the animation (movie 1) to Sirius and Hyades respectively, and L25 which presents a significantly smaller slope and is not referable back to the Solar radius. Finally, there are other structures with smaller extension in $R$, such as L1, L19 to 22 and L26, which show some continuity, meaning they are unlikely to be sporadic peaks. Yet they are present in a small range of $R$ and depict different shapes and trends compared to the prominent lines. 

\begin{figure*}
    \centering

\includegraphics[width=0.98\textwidth]{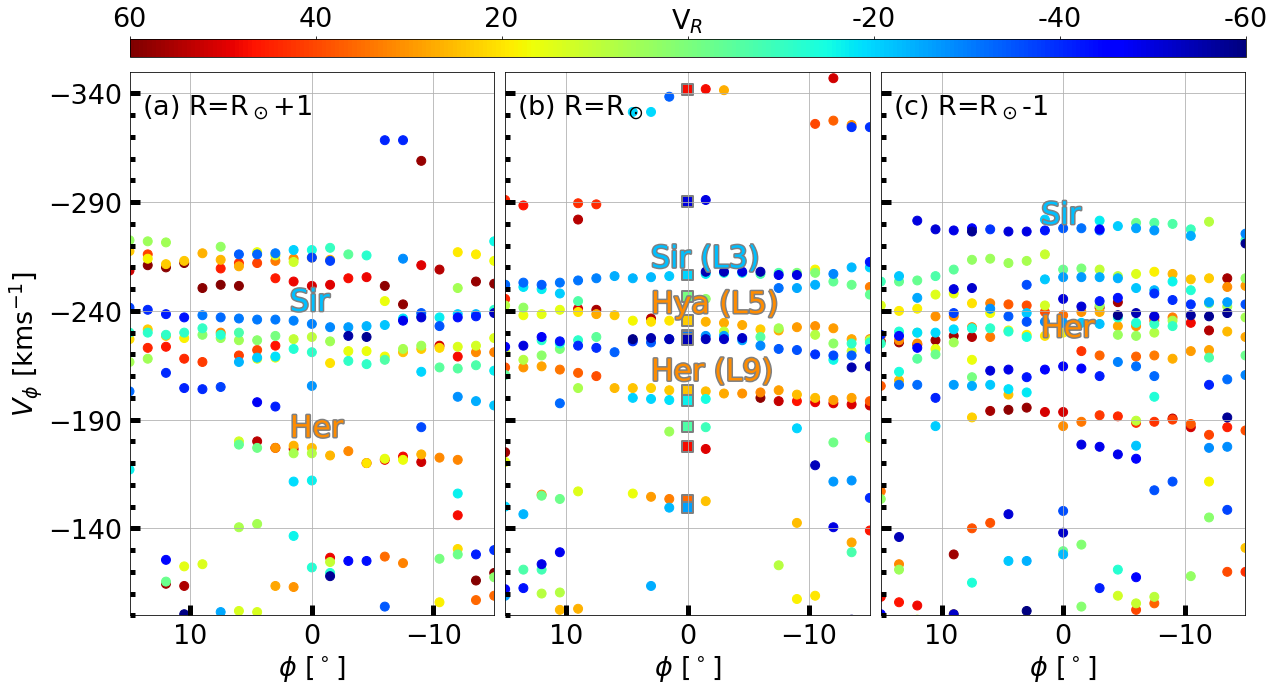} 

\caption{Evolution of the rotation velocity as a function of azimuth. Here, we show the peaks of the 63 volumes (21 at each radius) that fulfil the same conditions of Fig. \ref{fig:2exp_R}. The central panel corresponds to the exploration at the Solar radius, while in left panel we explore outer radius and on the right, we show the results for inner radius.}
\label{fig:2exp_P}
\end{figure*}

Changing to the variations along azimuth, Fig. \ref{fig:1exp_P} shows four different wavelet planes at scale $j=4$, corresponding to the volumes centred at $\phi$=10.5,4.5,-4.5,-10.5 deg. As in previous works \citep{Antoja2012}, we do not see much variation,  with only more subtle changes in the sub-structure compared to their evolution with $R$. We do observe a drift of Hercules, which moves to higher $|V_\phi|$ and becomes more prominent when increasing $\phi$. There is also a structure that at V13 appears as a single, almost isolated, peak around $V_\phi\sim$-230, $V_R\sim$-40 $\kms$, which corresponds to the mixture of Dehnen98-6 and Dehnen98-14, that for negative azimuth gains strength and forms an elongated structure. 

In an analogous way to Fig. \ref{fig:2exp_R}, in Fig. \ref{fig:2exp_P} we now plot $V_\phi$ against $\phi$ of the peaks found in each of the 21 different Galactic neighbourhoods that cover different azimuths $\phi$, for three different radius. Focusing on the middle panel, corresponding to the Solar radius, we confirm that most of the structures do not present significant variation with $\phi$, except for the L5 and L9 corresponding to Hyades and Hercules, respectively. The variation in azimuthal velocity in these cases is of about 13 $\kms$ in 30 deg, or equivalently,  $\sim$0.43 $\kms$ deg$^{-1}$, which at Solar radius means $\sim$3 $\kms$ kpc$^{-1}$, and thus much smaller than the average 23 $\kms$ kpc$^{-1}$ in the radial direction. As for the variations in radial velocity revealed by the colours, we see a slight shift of the peak towards larger positive $V_R$ at $\phi$=7.5 deg for Hercules.
In contrast, Sirius (L3) remains rather unchanged.

In the left and right panels of Fig. \ref{fig:2exp_P}, we show the azimuthal evolution in two different radius, where we basically see all peaks moving to larger or smaller $V_\phi$, in agreement with previous figures. In panel (a), corresponding to outer radius, at $V_\phi\sim$-175 $\kms$, the slope of Hercules with azimuth is still appreciable. In the same panel, among all the peaks at $\phi\geq$7.5$^\circ$, we distinguish a new feature at $\sim$-205 $\kms$ worth investigating since it seems related to Dehnen98-14 when inspecting the animations.


\section{Discussion}
\label{sec:discussion}

Although other studies have proposed links of rounded kinematic groups in the shape of a few diagonal branches \citep{Skuljan1999,Antoja2008} and ripple-like structures \citep{Xia2015}, the arched distribution depicted by the WT of the \textit{Gaia} data is radically distinctive and clear.

\paragraph{Arches of constant energy.} We have seen that some arches appearing in the Solar Neighbourhood velocity distribution have the same kinetic energy at all its points in the velocity plane. This matches the models of \cite{Minchev2009} and \citet{Gomez2012a} who showed how a phase mixing process after a perturbation of the potential, possibly due to a satellite passage, naturally leads to arches in the velocity plane of constant energy. 
Instead, some other arches that we detect do not follow lines of constant energy and are not symmetric in their Galactocentric radial velocity. In these lines, \citet{Antoja2009} showed with simulations that a barred potential acting on a disk suffering from phase mixing will make the arches deviate from being symmetric. Also, a non-axisymmetric potential (i.e., with bar and/or spiral arms) by itself can create similar arches not centred with respect to the Galactocentric radial velocity. For instance, the studies by \citet{Fux2001,Quillen2005,AntojaPhD,MonariPhD,Hattori2018} demonstrated that periodic orbits in the velocity plane have mostly an elongated locus with a peculiar arched shape. Nevertheless, while these are mostly theoretical studies analysing the regularity of the orbits, simulations of orbit integrations with realistic phase space distributions fail to populate distinctively many of these arches with particles \citep[e.g.,][]{Antoja2011,Hattori2018}, especially for spiral arms. Instead, some simulations with bars do show few but clear overdensities (and gaps) in the shape of arches \citep[e.g., ][]{Dehnen2000,Fux2001}.  


\paragraph{Decrease of $|V_\phi|$ with $R$.} We also note that most of the the peaks detected by the WT decrease their azimuthal velocity with Galactocentric radius, i.e., have a negative $\frac{\partial |V_\phi|}{\partial R}$. 
Since the peaks that we detect are overdensities in phase space, their change in $R$ is consistent with the diagonal ridges found in the $R$-$V_\phi$ projection of the recent study by  \citet{Antoja2018} (their Fig. 3).
Moreover, we find that the change in $R$ of some of the of the peaks such as Hercules and Hyades is consistent with being structures of roughly constant angular momentum ($L_Z$), i.e. follow lines of constant $L_Z$ in the $R$-$V_\phi$ plane, in contrast to, for instance, Sirius. This is to be expected in a resonance \citep{Sellwood2010,Quillen2018} since resonant features are composed of stars following similar orbits (roughly the same eccentricity) but at different amplitudes and phases. Given that not all of the arches found in the velocity plane of the Solar Neighbourhood evolve in the same manner, this could indicate that some are due to resonances and some to a phase mixing event. In turn, this would reduce the number of phase mixing "waves" and, thus, increase their velocity separation. A larger separation would reduce the estimated time of perturbation of $\sim$2 Gyr in \cite{Minchev2009}, potentially reconciling their assessment and the one by \cite{Antoja2018} from the curling of the ``snail shell'' in the vertical projection of phase space.


\paragraph{The Hercules stream.}Other authors have observed a correlation between the azimuthal velocity of Hercules and $R$, which was used in \cite{Antoja2014} and \cite{Monari2017} to constrain the pattern speed of the bar. Our analysis returns a slope in the $V_\phi-R$ plane for Hercules of 26.5$\pm$0.2 kms$^{-1}$kpc$^{-1}$, compatible with those studies. Nonetheless, we traced this trend for $\sim$2 kpc as opposed to the $\sim$0.6 kpc from previous studies. Additionally, we observe for the first time changes in the radial velocity $V_R$ of Hercules, which varies from $\sim$0 $\kms$ ($R\sim7.5$ kpc) to $\sim$40 $\kms$ ($R\sim9$ kpc). 
This gradient is in agreement with the behaviour predicted by \cite{Dehnen2000}, where a structure caused by the Outer Lindblad Resonance (OLR) of the bar evolves from zero radial velocity, towards smaller |$V_\phi$| and positive $V_R$ with increasing radius. This model also predicts an increase of the azimuthal velocity with $\phi$, which we also observe in the \textit{Gaia} data. Therefore, assuming that the gap between L5 and L9 in Fig. \ref{fig:2exp_R} is caused by the OLR, a direct application of equation (6) from \cite{Quillen2018} using the average $L_Z$ of Hercules and Hyades yields a rough estimate of the bar pattern speed of $\Omega_b\sim$54 $\kms$kpc$^{-1}$.
In contrast, recently \cite{Hunt2018b} explained the Hercules moving group with the 4:1 OLR of a slow bar. 
We note that the elongated feature in their Fig.7 could also explain some of the arches that we found in the upper part of the velocity distribution. In parallel, \cite{Hattori2018} explored different models and, to our knowledge, the one that reproduces better the curvature of the gap above Hercules is their fast-bar-only. 
Moreover, their exploration of where in the disk the bi-modality is observed (their Fig. 10) is more consistent with our azimuthal exploration in the case of their fast-bar-only model: both data and model show an increase of strength in Hercules with positive azimuths.
\cite{Bienayme2018} also studied the evolution of Hercules at different Galactic neighbourhoods, but in this case our range of exploration does not allow for a comparison of the key features in his simulations. In any case, further study of the Hercules structure is necessary to settle the debate about its origin and the controversy of the bar being slow or fast, first explored by \citep{PerezVillegas2017}.



\paragraph{Other kinematic sub-structures.} Another structure with unclear origin is Arcturus, which has been linked to accreted stars from a satellite Galaxy \citep{Navarro2004} but also to a resonance of the bar \citep{Williams2009,Antoja2009}. Here we observe that Arcturus evolves with $R$ similarly to Hercules, pointing to a dynamical origin. We also observed other peaks that show a distinct evolution with $R$ compared to the main groups, i.e. flat, oscillating or even increasing $V_\phi$. Most of these have low angular momentum and thus could have a different origin, perhaps related to accreted structures or groups in the nearby halo \citep{Helmi1999b,Koppelman2018}, although their significance needs to be explored further.



\paragraph{Improvements over previous work.} 
Our spatial exploration is inspired by, but unquestionably improves over, previous ones such as \cite{Antoja2012,Antoja2014}, who focused mostly on Hercules and for a short range of $R$, and \cite{Quillen2011}, that explored mostly changes in Galactic longitude but at very close distances. In comparison with other analysis \citep[e.g.,][]{Monari2017,Antoja2018,Kawata2018}, we decompose each volume independently with the WT which grants us the sensitivity necessary to explore the farthest neighbourhoods and allows us detect more and better defined ridges. 
Also, it introduces a third variable ($V_R$), which in the case of \cite{Monari2017} is done through cuts in the data at the cost of reducing the number of stars.


\section{Conclusions}
\label{sec:conclusions}

We have studied the velocity plane in Cylindrical coordinates for a \textit{Gaia} sample of stars with 6D phase space coordinates both in the Solar Neighbourhood, with 435,801 stars, and other Galactic neighbourhoods up to Heliocentric distances of $\sim$ 2.5 kpc. With the unprecedented number of stars and the small uncertainties in the observables, with median velocity errors of $\tilde{\sigma}_{vel} \leq 1$ $\kms$ in the vicinity of the Sun 
and $\leq 3$ $\kms$ for the rest of volumes, the kinematic sub-structures unveiled by the WT show a remarkable sharpness. We have characterized these structures, and followed their evolution with Galactocentric radius and azimuth. 

The WT reveals conspicuous thin kinematic arches with various morphologies that were not observed prior to \textit{Gaia} due to resolution limitations. We find at least 6 long arches in the Solar Neighbourhood with a nearly constant azimuthal velocity and wide range of radial Galactocentric velocities, plus 6 additional shorter segments of arch. Based on the scale in the WT where they are better detected, we find that they have a width of 2-8 $\kms$.  
Thanks to the superb isotropy of the \textit{Gaia} survey and the capabilities of the WT for unveiling structures, the exploration of the sub-structure in the velocity distribution can now be carried out in many different Galactic neighbourhoods, that is in all directions and spanning an astonishing range of 5 kpc in Galactocentric distance. In our analysis, we see that the azimuthal velocity of the peaks detected changes strongly with Galactocentric distance at an average rate of $\sim$23 $\kms$kpc$^{-1}$. We also see clear differences in their radial range, as some structures rapidly fade, while others can be identified at Galactic neighbourhoods 3 kpc apart. 
When explored as a function of Galactic azimuthal angle, most of the sub-structure stay constant and some present little change of the order of 3 kms$^{-1}$kpc$^{-1}$, thus an order of magnitude smaller than the changes in Galactic radius. This is qualitatively in agreement with expectations from models of resonant structures \citep{Antoja2011,Quillen2011} and of phase mixing \citep{Gomez2012a}.
Overall, when looking at their evolution in the velocity plane, most of these structures shift in the vertical axis ($V_\phi$) at a similar rate, producing the illusion that we are indeed riding kinematic waves in the Milky Way disk. 

With the fine characterization of the arches and ridges done in this work, 
we have seen that some of the kinematic sub-structure follow lines of constant energy at a given volume and could be related to phase mixing processes \citep{Minchev2009,Gomez2012a} induced, e.g., by the close passage of the satellite dwarf. These coincide with the structures showing less variations in azimuthal angle, like Sirius. Others, such as Hyades or Hercules, seem to follow lines of constant angular momentum, similar to what is expected in case of resonant kinematic sub-structure \citep{Sellwood2010,Quillen2018} induced by the bar and/or the spiral arms of our Galaxy, and show also variations with azimuthal angle. 
We could be, thus, witnessing several mechanisms acting on the disk, as recently suggested in \cite{Antoja2018}, who compared the ridges with different toy models of phase mixing and resonances, and in \cite{Trick2018} in their study of the \textit{Gaia} data in action space.  Here we provide additional evidence of this possible duality and a tentative way to differentiate between structures of different origin.

We have also shown the evolution of $V_\phi$ for Hercules both with radius and azimuth, reaching for the first time distances of $\sim$2 kpc. Its changes with radius are consistent with models of Hercules produced by the OLR of the Galactic bar, but also we have, for the first time, quantified the changes in azimuth and of its radial velocity, which still remain consistent with the model. Our exploration has lead also to the first unambiguous extrasolar kinematic structures, that is moving groups and arches not observed (or very weak) at Solar radius. Some of these new sub-structures follow similar trends with Galactocentric radius and azimuth as the sub-structure crossing the local volume, whereas others evolve distinctively. 

With this work, and those to come, a new window for the modelling of our Galaxy opens. 
Future efforts should be devoted to explain the various structures found and their evolution in the context of coexisting and perhaps interacting processes, since we have now the observational power to study -and contrast- the effects of different dynamical mechanisms. 





\begin{acknowledgements}\\
This work has made use of data from the European Space Agency (ESA)
mission {\it Gaia} (\url{https://www.cosmos.esa.int/gaia}), processed by
the {\it Gaia} Data Processing and Analysis Consortium (DPAC,
\url{https://www.cosmos.esa.int/web/gaia/dpac/consortium}). Funding
for the DPAC has been provided by national institutions, in particular
the institutions participating in the {\it Gaia} Multilateral Agreement.
This project has received funding from the University of Barcelona's official doctoral program for the development of a R+D+i project under the APIF grant and 
from the European Union's Horizon 2020 research and innovation programme under the Marie Sk{\l}odowska-Curie grant agreement No. 745617. 
This work was supported by the 
MINECO (Spanish Ministry of Economy) through grants ESP2016-80079-C2-1-R (MINECO/FEDER, UE) and ESP2014-55996-C2-1-R (MINECO/FEDER, UE). 

\end{acknowledgements}


\bibliographystyle{aa}
\bibliography{Biblio}

\appendix

\section{The proper use of parallaxes}
\label{App:parallax}
Here we discuss briefly the recommendations given on the use of the \textit{Gaia} parallaxes. For a more detailed study, we refer the reader to \cite{Luri2018}. The problem of inverting the parallax is that it generates an asymmetric and biased distribution of probability for the true distance given the evidence (observations). In other words, the most probable distance is not the \textbf{true} distance of the star. Therefore, using summary statistics like, for instance, the mode or the mean causes a bias in the distances derived for each star. Moreover, the simple $1/\varpi$, which in a Bayesian context is equivalent to select the most likely value given a uniform \textit{prior}, has been shown to be a quite bad estimator \citep{Brown1997,Arenou1999,DR2-DPACP-38}. In addition, as shown in \cite{Luri2018}, cutting in parallax relative error, $\sigma_\varpi/\varpi$ is actually far from beneficial. It can introduce an even deeper bias depending on the characteristics of the sample. In contrast, the recommendation given by the \textit{Gaia} team is to work fully Bayesian, that is, do not use summary statistics and, instead, drag the \textit{Posterior} probability distributions up to the quantities of interest. The results will then be, not a value, but a whole p.d.f. filled with information. In general, though, it is enough to use the simple and biased estimator 1/$\varpi$ to explore the data, always taking into account the effects it can introduce if the sample is not very close to the Sun and has large uncertainties.

Our sample, as described in Sect. \ref{sec:sample}, is formed mainly by bright nearby stars, precisely those with the smallest mean uncertainties in the DR2 catalogue. Moreover, for nearby stars with fractional error smaller than 0.2, the dependence on the \textit{prior} is highly attenuated and the bias of using the inverse of the parallax as a distance estimator becomes negligible (e.g., \citealt{Bailer-Jones2015,Astraatmadja2016}). Beyond that, our science case and conclusions do not rely heavily on the distance determination, and the potentially small bias does not affect the overall picture of the sub-structure in the kinematic plane. 

As a side note, we notice that by using the Jacobian to propagate and calculate the uncertainties in the velocities, we are implicitly modelling the errors as Gaussian which is known to be false. As already mentioned, the inversion of the parallax results in an asymmetric probability distribution, also for the observed $V_R$ and $V_\phi$. 
We do not impose any threshold or sigma-clipping based on this errors. In this sense, since we already selected only stars with "good" parallax, the assumption of Normality is unimportant.

\section{On-line material}  
\label{App:material}
The on-line material referenced in this work is available at \href{https://sites.google.com/fqa.ub.edu/pramos}{https://sites.google.com/fqa.ub.edu/pramos}.

\begin{itemize}
\item Animation 1: WT at the scale 8-16 $\kms$ of the velocity plane at different Galactocentric distances.
\item Animation 2: Same as animation 1 but at the scale 4-8 $\kms$.
\item Animation 3: Same as animation 1 but at the scale 16-32 $\kms$.
\item Animation 4: All three scales ($j$=3,4,5) evolving simultaneously with Galactocentric distance at the Sun-GC line.
\item Animation 5: All three scales ($j$=3,4,5) evolving simultaneously with azimuth at Solar radius.
\item Table \ref{App:material}.1: All the significant peaks in the Solar Neighbourhood sample at the scale 4-8 $\kms$ ($j$=3).
\end{itemize}

\section{List of Moving Groups}  
\label{App:MG}
We compiled a list of moving groups from the corresponding tables of \cite{Dehnen1998,Antoja2012,Xia2015,Bobylev2016,Liang2017,Kushniruk2017} and list them in Table \ref{tab:all_mg}. 
We transformed their velocities into Galactocentric Cylindrical coordinates following the recipe given in Sect. \ref{sec:sample}. Some moving groups, like Hyades or Sirius, have more than one entry since each author derived independently a different location in velocity space. 

For \cite{Kushniruk2017} we only took the new unnamed objects (at scales 3 and 4, see \citealt{Kushniruk2017}) in order to reduce the scatter in our table since in their study the same structure appears several times in their tables.
Their newly found objects are label here including the scale they were found (since their notation and definition of the scales coincides with the one we used for this work). 
Regarding the elongated structures described by \cite{Xia2015}, where just the start and end points are given, we did not include them to avoid miss-matches, but we considered them for the discussion. 
The rest of new or unnamed objects found in each author's table are labelled as \textbf{\{First Author\} + \{Year\} + \{Number in source table\}}. 

The names of the objects in our Table \ref{tab:scale4} are given according to the following criteria: first, we cross-matched the position of the peak detected in our data with the coordinates in the literature using a circle of radius $\Delta$2$^j$ $\kms$. Obviously, since we may have several entries for a single object, we grouped the candidates by name and counted their frequency. Then, in the case of obtaining multiple groups, we selected the one with more counts. If we had a draw, we selected the closest one to the peak.

\onecolumn
\begin{longtable}{lllrrr}
\caption{\label{tab:all_mg} Compilation of moving groups from different authors (see text). First column contains the author and the year of publication, whereas the second one shows the name of the object and the fourth, the corresponding number in the original table. The source of the astrometry is shown in column three. Finally, the cylindrical velocities (columns 5 and 6) are derived using the parameters for the Sun and LSR described in Sect. \ref{sec:sample}.}\\
\hline\hline
Author & Name & Source & Original number & $V_R$ [$\kms$]& $V_\phi$ [$\kms$]\\
\hline
Antoja2012	&	Antoja12-12	&	RAVE	&	12	&	-103	&	-229	\\
Antoja2012	&	Antoja12-13	&	RAVE	&	13	&	92	&	-211	\\
Antoja2012	&	Antoja12-GCSIII-13	&	GCSIII	&	-	&	69	&	-260	\\
Antoja2012	&	Antoja12-15	&	RAVE	&	15	&	-71	&	-180	\\
Antoja2012	&	Antoja12-16	&	RAVE	&	16	&	111	&	-233	\\
Antoja2012	&	Antoja12-17	&	RAVE	&	17	&	-23	&	-144	\\
Antoja2012	&	Antoja12-18	&	RAVE	&	18	&	-120	&	-220	\\
Antoja2012	&	Antoja12-19	&	RAVE	&	19	&	-38	&	-132	\\
Dehnen1998	&	Arcturus	&	Hipparcos	&	10	&	-11	&	-142	\\
Xia2015	&	Arifyanto05	&	LAMOST-RAVE	&	8	&	53	&	-172	\\
Bobylev2016	&	Bobylev16-14	&	RAVE DR4	&	14	&	-41	&	-212	\\
Bobylev2016	&	Bobylev16-15	&	RAVE DR4	&	15	&	85	&	-242	\\
Antoja2012	&	Coma Berenices	&	RAVE	&	1	&	-4	&	-246	\\
Bobylev2016	&	Coma Berenices	&	RAVE DR4	&	1	&	-4	&	-246	\\
Dehnen1998	&	Coma Berenices	&	Hipparcos	&	4	&	-1	&	-247	\\
Xia2015	&	Coma Berenices	&	LAMOST-RAVE	&	4	&	0	&	-245	\\
Dehnen1998	&	Dehnen98-11	&	Hipparcos	&	11	&	59	&	-242	\\
Dehnen1998	&	Dehnen98-12	&	Hipparcos	&	12	&	59	&	-202	\\
Dehnen1998	&	Dehnen98-13	&	Hipparcos	&	13	&	-61	&	-252	\\
Liang2017	&	Dehnen98-14	&	LAMOST-TGAS	&	5	&	-62	&	-224	\\
Dehnen1998	&	Dehnen98-14	&	Hipparcos	&	14	&	-61	&	-227	\\
Antoja2012	&	Dehnen98-14	&	RAVE	&	7	&	-59	&	-228	\\
Bobylev2016	&	Dehnen98-14	&	RAVE DR4	&	7	&	-54	&	-228	\\
Dehnen1998	&	Dehnen98-6	&	Hipparcos	&	6	&	-31	&	-232	\\
Bobylev2016	&	Dehnen98-6	&	RAVE DR4	&	12	&	-31	&	-230	\\
Dehnen1998	&	Dehnen98-8	&	Hipparcos	&	8	&	29	&	-202	\\
Dehnen1998	&	Dehnen98-9	&	Hipparcos	&	9	&	14	&	-202	\\
Antoja2012	&	$\epsilon$Ind	&	RAVE	&	10	&	70	&	-210	\\
Liang2017	&	$\epsilon$Ind	&	LAMOST-TGAS	&	8	&	72	&	-213	\\
Bobylev2016	&	$\epsilon$Ind	&	RAVE DR4	&	10	&	79	&	-203	\\
Antoja2012	&	$\eta$Cep	&	RAVE	&	14	&	16	&	-157	\\
Bobylev2016	&	$\eta$Cep	&	RAVE DR4	&	11	&	31	&	-151	\\
Liang2017	&	$\gamma$Leo	&	LAMOST-TGAS	&	10	&	-81	&	-246	\\
Antoja2012	&	$\gamma$Leo	&	RAVE	&	11	&	-79	&	-253	\\
Bobylev2016	&	$\gamma$Leo	&	RAVE DR4	&	9	&	-76	&	-253	\\
Antoja2012	&	$\gamma$Leo	&	RAVE	&	9	&	-67	&	-254	\\
Liang2017	&	$\gamma$Leo	&	LAMOST-TGAS	&	7	&	-61	&	-255	\\
Liang2017	&	Hercules I	&	LAMOST-TGAS	&	6	&	41	&	-211	\\
Antoja2012	&	Hercules I	&	RAVE	&	6	&	46	&	-204	\\
Bobylev2016	&	Hercules I	&	RAVE DR4	&	8	&	46	&	-204	\\
Liang2017	&	Hercules II	&	LAMOST-TGAS	&	4	&	17	&	-204	\\
Antoja2012	&	Hercules II	&	RAVE	&	8	&	17	&	-202	\\
Bobylev2016	&	Hercules II	&	RAVE DR4	&	5	&	24	&	-202	\\
Liang2017	&	HR1614	&	LAMOST-TGAS	&	11	&	-32	&	-188	\\
Dehnen1998	&	HR1614	&	Hipparcos	&	7	&	-26	&	-192	\\
Bobylev2016	&	HR1614	&	RAVE DR4	&	13	&	-25	&	-194	\\
Antoja2012	&	Hyades	&	RAVE	&	2	&	19	&	-239	\\
Bobylev2016	&	Hyades	&	RAVE DR4	&	2	&	19	&	-237	\\
Liang2017	&	Hyades	&	LAMOST-TGAS	&	0	&	19	&	-236	\\
Dehnen1998	&	Hyades	&	Hipparcos	&	2	&	29	&	-232	\\
Xia2015	&	Hyades-Pleiades	&	LAMOST-RAVE	&	5	&	7	&	-234	\\
Kushniruk2017	&	Kushniruk17-J4-2	&	RAVE-TGAS	&	2	&	-49	&	-259	\\
Kushniruk2017	&	Kushniruk17-J3-13	&	RAVE-TGAS	&	13	&	-48	&	-260	\\
Kushniruk2017	&	Kushniruk17-J3-18	&	RAVE-TGAS	&	18	&	75	&	-176	\\
Kushniruk2017	&	Kushniruk17-J3-19	&	RAVE-TGAS	&	19	&	7	&	-185	\\
Liang2017	&	Liang17-12	&	LAMOST-TGAS	&	12	&	11	&	-185	\\
Liang2017	&	Liang17-13	&	LAMOST-TGAS	&	13	&	41	&	-265	\\
Liang2017	&	Liang17-14	&	LAMOST-TGAS	&	14	&	25	&	-281	\\
Liang2017	&	Liang17-9	&	LAMOST-TGAS	&	9	&	-19	&	-201	\\
Dehnen1998	&	NGC1901	&	Hipparcos	&	5	&	14	&	-242	\\
Liang2017	&	Pleiades	&	LAMOST-TGAS	&	1	&	-5	&	-229	\\
Dehnen1998	&	Pleiades	&	Hipparcos	&	1	&	1	&	-230	\\
Bobylev2016	&	Pleiades	&	RAVE DR4	&	4	&	2	&	-228	\\
Antoja2012	&	Pleiades	&	RAVE	&	4	&	5	&	-230	\\
Liang2017	&	Sirius	&	LAMOST-TGAS	&	2	&	-27	&	-256	\\
Xia2015	&	Sirius	&	LAMOST-RAVE	&	3	&	-22	&	-251	\\
Bobylev2016	&	Sirius	&	RAVE DR4	&	3	&	-21	&	-253	\\
Dehnen1998	&	Sirius	&	Hipparcos	&	3	&	-20	&	-255	\\
Antoja2012	&	Sirius	&	RAVE	&	3	&	-15	&	-256	\\
Bobylev2016	&	Wolf 630	&	RAVE DR4	&	6	&	-40	&	-231	\\
Antoja2012	&	Wolf 630	&	RAVE	&	5	&	-39	&	-231	\\
Liang2017	&	Wolf 630	&	LAMOST-TGAS	&	3	&	-34	&	-230	\\

\hline
\hline
\end{longtable}
\twocolumn

\end{document}